# Observation of Macroscopic Nonlocal Voltage at Room Temperature

Jae Ho Jeon[1]*, Hong Ryeol Na[1], Sahng-Kyoon Jerng[1], Seyoung Kwon[2], Sungkyun Park[2], Kang Rok Choe[3], Jun Sung Kim[3], Heeju Kim[4,5], Gunn Kim[5], Sangmin Ji[6,7], Taegeun Yoon[6,7], Young Jae Song[6,7,8,9], Dirk Wulferding[1], Jeong Kim[10], Hwayong Noh[1], and Seung-Hyun Chun[1]*

[1]Department of Physics, Sejong University, Seoul 05006, Republic of Korea

[2]Department of Physics, Pusan National University; Busan, 46241, Republic of Korea.

[3]Department of Physics, Pohang University of Science and Technology; Pohang, 37673, Republic of Korea.

[4]Department of Physics and HMC, Sejong University, Seoul 05006, Republic of Korea.

[5]School of Computational Sciences, Korea Institute for Advanced Study, Seoul, 02455, Republic of Korea.

[6]SKKU Advanced Institute of Nano Technology (SAINT), Sungkyunkwan University (SKKU); Suwon, 16419, Republic of Korea,

[7]Department of Nano Science and Technology, Sungkyunkwan University (SKKU); Suwon, 16419, Republic of Korea.

[8]Department of Nano Engineering, Sungkyunkwan University (SKKU); Suwon, 16419, Republic of Korea,

[9]Center for 2D Quantum Heterostructures, Institute for Basic Science, Department of Physics, Sungkyunkwan University (SKKU); Suwon, 16419, Republic of Korea.

[10]Department of Electrical Engineering, Sejong University, Seoul 05006, Republic of Korea

* To whom correspondence should be addressed. Email: jeonjh713@gmail.com and schun@sejong.ac.kr

**ABSTRACT**

Electrons in conductors suffer frequent scattering from defects and phonons, and their diffusive macroscopic behavior is governed by an external electric field through Ohm's law. If electrons are no longer diffusive, the Ohmic description collapses. In devices composed of thin chalcogenides and $YBa_2Cu_3O_{7-\delta}$, we observe a transition from an Ohmic conductor to a "nonlocal conductor" below a certain temperature. The "nonlocal conductor" is characterized by significant nonlocal voltages (~0.1 V) across current-free macroscopic regions (~1 mm) that are conventionally considered equipotential. Simultaneously, an equivalent excess voltage appears in the current-carrying region in addition to the Ohmic voltage. When the Ohmic resistance vanishes at low temperature, the potential profiles of the current-carrying and current-free regions become identical, confirming the intrinsic nature of the nonlocal voltages. Negative local resistances in a vicinity geometry support macroscopic hydrodynamic flow as the underlying mechanism, implying electron momentum conservation over unprecedented distances. This new conduction state, observable at room temperature, opens the field of nonlocal electronics and low-dissipation applications.

**KEYWORDS:** nonlocal voltage, nonlocal conductor, macroscopic, chalcogenides, hydrodynamics

At room temperature, the mean-free paths of ordinary metals and semiconductors are limited to ≤ 100 nm.[1,2] Even in encapsulated graphene, the cleanest material known so far, electrons cannot travel farther than 1 μm before being scattered.[3] Therefore, when considering millimeter-scale devices and electrical contact separations, local resistivity is a well-defined quantity for characterizing a material system. In the absence of temperature and electron-density gradients, the local resistivity is readily obtained from the potential difference between any two points along the current path and the device geometry.

We define "nonlocal voltage" as the potential difference measured at points outside the conventional current path, thereby contradicting the assumption of local resistivity. To date, nonlocal phenomena have been observed in mesoscopic systems and have been attributed to various quantum interactions, from spin diffusion to topological states of matter. Electrical spin injection has been verified through magnetic-field-dependent nonlocal voltages in metals, semiconductors, and Dirac materials.[4–7] In Hall-bar devices, nonlocal resistance can serve as evidence of edge-channel transport in the quantum spin Hall state [8] or as a manifestation of the spin Hall effect. Topological insulators (TIs) and Weyl semimetals have also exhibited similar nonlocal voltages.[9–12] However, room temperature observations have been limited to spin-diffusive phenomena, with lateral dimensions of less than 2 μm at best.[5,13]

While investigating the proximity effects between TIs and high-transition-temperature ($T_C$) superconductors (SCs), we observed macroscopic nonlocal voltages (MNVs) at distances exceeding 1 mm from the current path. For 20 nm-thick $Bi_2Se_3$ (BS) film grown on nominal $YBa_2Cu_{3-y}O_{7-\delta}$ (YBCO), the MNV measured at 70 μm from the current contact exceeded 400 mV at room temperature. Subsequent investigations revealed similar behavior in materials other than TIs, including $CrTe_x$ and $MnTe_x$ (MT) grown on YBCO. Not all, but most bilayers showed nonlinear current-voltage (*I-V*) characteristics in four-point-probe measurements. Because the observed MNV is unprecedented and cannot be explained by any previously reported nonlocal mechanism, we designate this state as a "nonlocal conductor".

In this paper, we introduce the "nonlocal conductor" as a new conduction state hosted within the chalcogenide layer. The state emerges below a characteristic temperature, $T_{NL}$, which depends on the chalcogenide composition. In the absence of an external electric field, the existence of MNV implies the presence of a potential gradient from viscous electron flow. The same voltage contribution also appears in the local configuration, far exceeding the conventional Ohmic voltage. The temperature dependence of position-dependent voltages clearly distinguishes the Ohmic and hydrodynamic contributions. This hydrodynamic interpretation is further supported by the observation of negative local resistances in a vicinity geometry. Although the microscopic origin of this phenomenon remains unclear, we believe that the "nonlocal conductor" challenges the conventional understanding of charge transport and may open the door to practical nonlocal electronics as well as a new strongly correlated quantum phase.

**Bilayer growth**

A bilayer system composed of SC and TI is currently attracting considerable interest because of its potential for realizing topological superconductivity and topological qubits. Following the theoretical proposal of Fu and Kane[14], numerous studies have investigated proximity-induced superconductivity in TIs.[15–17] Signatures of Majorana zero modes have reported in heterostructures composed of conventional layered SCs and TIs. [18,19] In contrast, analogous studies with high-$T_C$ SC have yielded controversial results.[20–24] Furthermore, TI-SC proximity studies involving YBCO remain relatively rare.[25] Motivated by this opportunity, we initiated the fabrication of BS/YBCO bilayers.

We used commercial 50 nm-thick YBCO films and grew 20 nm-thick BS by van der Waals epitaxy using molecular beam epitaxy (MBE). Building on our previous experience with BS growth on amorphous substrates, we employed a Se-passivation step prior to BS deposition[26]. This approach enabled us to overcome challenges arising from the crystal-symmetry mismatch between BS and YBCO. However, the resulting structures differed significantly from our original expectations. High-resolution transmission electron microscopy (HR-TEM) and energy-dispersive X-ray spectroscopy (EDS) revealed considerable Cu diffusion from the YBCO layer into the overlying BS layer despite the relatively low growth temperature.[27] The concentration of diffused Cu, estimated to be $x \approx 1$, significantly exceeded the thermodynamic solubility limit of homogeneous $Cu_xBi_2Se_3$ ($x = 0.3$) (Supplementary Fig. 1).[28] We note that Cu concentrations as high as $x = 7.5$ have previously been reported using solution-based and electrochemical intercalation methods.[29,30]

X-ray diffraction measurements exhibited broad diffraction peaks associated with both YBCO and BS. The grain sizes were estimated to be approximately 80 nm and 20 nm for YBCO and BS, respectively (Supplementary Fig. 2). Because the crystal structure of BS remained intact, as further confirmed by Raman spectroscopy (Supplementary Fig. 3), most Cu atoms were likely incorporated as interstitials.[29] Density functional theory (DFT) calculations suggested that Cu atoms can reside within the van der Waals gaps, forming tetragonal bonds with Se atoms while preserving the host crystal structure (Supplementary Fig. 4). Despite substantial Cu diffusion, the YBCO layer retained its superconducting properties, albeit with a reduced $T_C$ (Supplementary Fig. 5).

Following the discovery of MNV in BS/YBCO, we investigated additional chalcogenide materials and found that TI behavior is not a prerequisite for the emergence of MNV. Both $Bi_2Te_3$ (a TI) and $Sb_2Se_3$ (a non-TI) grown on YBCO exhibited clear MNVs. Furthermore, magnetic $CrTe_x$ and MT layers grown on YBCO also displayed pronounced MNVs when deposited at temperatures lower than those optimized for growth on insulating substrates. Unlike BS on YBCO, $CrTe_x$ and MT films were nanocrystalline and exhibited random grain orientations. Still, significant Cu migration from the YBCO layer was consistently observed in all cases (Supplementary Fig. 6).

**Macroscopic Nonlocal Voltage**

To establish the "nonlocal conductor" behavior independently of topological surface states, we first examine an MT/YBCO bilayer. Figures 1a and 1b show schematic illustrations of the "local" and "nonlocal" measurement configurations, a cross-section HR-TEM image, and an optical image of the sample with manually defined macroscopic contact lines. The six yellow contact pads in Fig. 1a correspond sequentially to the six silvery contact lines in Fig. 1b.

Under conventional transport conditions, the local voltage is expected to increase linearly with current, while the nonlocal voltage remains negligible. This expectation is confirmed by the 373 K data shown in Figs. 1c and 1d. At 200 K, however, the local *I-V* characteristics become highly nonlinear, as shown in Fig. 1c. Although nonlinear *I-V* responses have been reported in systems such as diodes, Schottky junctions, and space-charge limited conductors, those systems contain barriers or insulating regions between the voltage probes.[31] In contrast, the present system exhibits an apparent resistivity of only 1−10 mΩ cm, making such marked non-Ohmic behavior highly unusual.

More remarkably, Fig. 1d shows that a finite nonlocal voltage of approximately 0.1 V was detected at distances of approximately 1 mm from the current path at 200 K. We emphasize that the driving current is confined to the local region and that the spreading resistance becomes negligible at distances greater than 1 μm from the current contact.[32] Hence, no external electric field is expected in the nonlocal region. Nonetheless, the measured MNV was comparable in magnitude to the local voltage.

This observation constitutes the central experimental finding of the present work and is difficult to reconcile with the conventional understanding of charge transport. In addition to its macroscopic spatial extent and large magnitude, the MNV exhibits an extreme nonlinearity that distinguishes it from all previously reported nonlocal transport phenomena.[4–12,33] Accordingly, we identify this transport regime as a distinct conduction state, which we term a “nonlocal conductor”.

**Phase Transition**

For quantitative transport measurements, the bilayers were patterned to 200 μm-wide channels using standard photolithography (Supplementary Fig. 7a). Figure 2a shows the decrease in MNV with increasing temperature. The nonlocal voltage vanished completely at temperatures of 320 K and above, as shown in Fig. 2b. Concurrently the local voltage became linear in current, and the slope of the local *I-V* curve increased slightly with further temperature increase, consistent with conventional metallic behavior (Supplementary Fig. 8 and Fig. 2c). Quantitatively, the sheet resistance above $T_{NL}$ was approximately 200 Ω, consistent with that of Cu-deficient YBCO.

The simultaneous disappearance of the nonlocal voltage and recovery of an Ohmic local response provides strong evidence for a transition from a "nonlocal conductor" to an Ohmic conductor at $T_{NL}$ ~320 K in this sample. The nonlocal *I-V* curves could be scaled onto a single universal curve, as shown in Supplementary Fig. 9. The critical exponent of the order parameter (MNV in the present case), $\beta$, was approximately 3, a value that does not correspond to any known universality class.[34] This unusual exponent is consistent with the behavior shown in Fig. 2b, where the MNV approaches zero much more gradually than in conventional second-order phase transitions.

The transition temperature depends on the chalcogenide material deposited on YBCO. Indeed, $T_{NL}$ ~320 K observed in MT/YBCO was the lowest among the samples studied here. In BS/YBCO, both the local and the nonlocal voltages decreased substantially with increasing temperature (Fig. 2d). However, even at 373 K, the MNV remained greater than half of its low-temperature maximum value. At higher temperatures, the chalcogenide layer underwent irreversible degradation, most likely because of the loss of volatile chalcogen species. Thus, $T_{NL}$ could not be determined experimentally for BS/YBCO. On the other hand, BS/YBCO appears to be the most promising bilayer system for room-temperature studies and potential technological applications.

**Hydrodynamics and Two-Current Model**

Figures 2b and 2c show that the emergence of nonlocal voltage below $T_{NL}$ is accompanied by a drastic increase in the local voltage. This observation motivates a detailed investigation on the potential profiles in both the local and nonlocal configurations as functions of the distance $|x|$ from the current source. To this end, we employed a long-channel BS/YBCO device at room temperature (Supplementary Fig. 7b). At large distances from the current source contact ($V_4$ and $V_1$ in Fig. 3a), the nonlocal voltage vanished, and the local voltage exhibited conventional Ohmic behavior. In contrast, at short distances ($V_3$ and $V_2$ in Fig. 3a), a large nonlocal voltage exceeding 100 mV at 300 μA was observed for an electrode spacing of 70 μm (Fig. 3b and 3c). At the same distance, the local voltage ($V_2$) increased by an amount equal to the nonlocal voltage ($V_3$), exceeding the Ohmic contribution ($V_1$) by more than a factor of 5. The enhancement of the local voltage cannot be attributed to sample inhomogeneity, because the same behavior was reproduced after lateral displacement of the contact pads. Conversely, $V_1$ could be increased and rendered nonlinear simply by moving the current contacts closely.

The systematic evolution of the local voltage with distance from the current source provides further evidence that a local resistivity cannot be meaningfully defined in this system. It also rules out non-uniform current distribution or other forms of sample inhomogeneity as the origin of MNV. Specifically, the nonlocal potential ($\phi_{NL}$) and the local potential ($\phi_L$) become identical after subtraction of the linear Ohmic contribution ($\phi_{Ohmic}$), as shown in Figs. 3d and 3e. This observation strongly suggests that the emergence of MNV and the enhancement of the local voltage originate from the same physical mechanism.

The distinction between Ohmic and non-Ohmic contributions becomes even clearer when their temperature dependences are examined. Using a second BS/YBCO device, we measured local and nonlocal voltages down to 10 K. The superconducting transition is clearly visible in the temperature dependence of the local voltages ($V_1$ and $V_2$), whereas the nonlocal voltages ($V_3$ and $V_4$) increase as the temperature decreases. We note that the magnetic-field dependence remained negligible for perpendicular magnetic fields up to 1 T over the entire temperature range investigated, indicating that the observed phenomena are unlikely to originate from spin-related interactions (Supplementary Fig. 10).

The temperature-dependent potential profiles reveal the essential characteristics of the nonlocal conductor. In a conventional Ohmic conductor, the potential increases linearly in the current-carrying region ($x < 0$) and remains constant in the current-free region ($x > 0$). Separately prepared BS and YBCO single layers exhibit precisely such behavior (Supplementary Fig. 11). In contrast, the potential profile of the nonlocal conductor is characterized by a superimposed symmetric peak at the current source (Fig. 4d). The peak becomes more pronounced when the YBCO layer enters the superconducting state below 50 K, causing the Ohmic contribution to vanish. Under such conditions, all externally supplied current is expected to flow through the superconducting YBCO layer without dissipation, and no measurable voltage should appear. However, we observe exponentially decaying potential profiles in both the current-carrying and current-free regions, with amplitudes exceeding those observed at room temperature.

These observations suggest that the nonlocal phase is hosted within the chalcogenide layer. Additional support for this interpretation comes from the disappearance of MNV following selective notching of the BS layer (Supplementary Fig. 12). In this picture, the role of YBCO is to supply Cu atoms to the chalcogenide layer during growth and to share the externally applied current according to the relative sheet conductances of the two layers. Diffusive electron transport alone cannot account for the observed MNV because no driving force exists in the nonlocal region. Ballistic electrons could in principle propagate into both local and nonlocal regions; however, such transport would be nearly dissipationless except near interfaces and therefore cannot explain the large local voltages observed experimentally.[35] We therefore consider a hydrodynamic description of charge transport. Analogous to fluid flow from a localized source, viscous electrons may propagate into both local and nonlocal regions. We assume that the nonlocal phase supports such viscous electron flow.

To describe the current distribution and the resulting potential profiles, we employ a two-current model analogous to the two-fluid model of superconductivity.[36] Let $\boldsymbol{J}_O$ and $\boldsymbol{J}_v$ denote the Ohmic and viscous current densities, respectively. The linearized Navier-Stokes equation for a two-dimensional electron fluid can then be written as [37,38]

$$\sigma_i \nabla\phi_i(r) - D_i^2 \nabla^2 \boldsymbol{J}_i(\boldsymbol{r}) = -\boldsymbol{J}_i(r)\,, \qquad (1)$$

where $\sigma_i$ is the Drude conductivity, $\phi_i$ is the potential, $D_i$ is the characteristic length scale defined as $D_i \equiv \sqrt{\nu_i \tau_i}$ ($\nu_i$ is the kinematic viscosity and $\tau_i$ is the momentum relaxing scattering time), and the subscript $i$ is $O$ (Ohmic) or ν (viscous). For the Ohmic current, the Drude conductivity is replaced by the effective conductivity of BS/YBCO bilayer ($\sigma_O = (\sigma_{BS}^{sh} + \sigma_{YBCO}^{sh})/t$, $\sigma_{BS}^{sh}$ and $\sigma_{YBCO}^{sh}$ are the sheet conductances of BS and YBCO, respectively, and $t$ is the bilayer thickness). As the Ohmic current is irrotational and uniform across the cross-section, the potential in the local region satisfies

$$\nabla\phi_L(r) = \nabla\phi_O + \nabla\phi_v(r) = -\boldsymbol{J}_O/\sigma_O + D_v^2 \nabla^2 \boldsymbol{J}_v(\boldsymbol{r})/\sigma_v - \boldsymbol{J}_v(\boldsymbol{r})/\sigma_v \qquad (2)$$

In the nonlocal region, the Ohmic current is absent :

$$\nabla\phi_{NL}(r) = \nabla\phi_v(r) = D_v^2 \nabla^2 \boldsymbol{J}_v(\boldsymbol{r})/\sigma_v - \boldsymbol{J}_v(\boldsymbol{r})/\sigma_v \qquad (3)$$

As the viscous flow is symmetric with respect to the current-injection point, the model naturally explains the overlap of the local and nonlocal potential profiles after subtraction of the Ohmic contribution (Fig. 3e). When the YBCO layer becomes superconducting and $\sigma_O$ diverges, both the local and nonlocal voltages arise exclusively from the viscous current component (Fig. 4a). Under these conditions, the local and nonlocal potential profiles become nearly indistinguishable, apart from a small offset associated with the negative potential peak centered at the current sink.

Figures 3e and 4d further show that the voltage generated by viscous flow decays exponentially with distance from the source, enabling extraction of the characteristic length scale $D_v$. Because $D_v$ corresponds to the diffusion length associated with $\nabla \times \boldsymbol{J}_v(\boldsymbol{r})$, it also represents the decay length of the hydrodynamic signal.[3] Across dozens of BS/YBCO samples fabricated using different MBE systems, clear MNVs were consistently observed, although the detailed responses varied between growth batches. For quantitative comparison, only patterned devices were analyzed, and $D_v$ was compared with the magnitude of MNV (Supplementary Fig. 13). At room temperature, $D_v$ in BS/YBCO ranged from 0.12 mm to 1.1 mm while the MNV varied by nearly two orders of magnitude. This broad distribution likely reflects variations in Cu migration, which are more difficult to control than conventional chemical doping. MNVs were also observed in $Bi_2Te_3$/YBCO, $Sb_2Se_3$/YBCO, and $CrTe_x$/YBCO. Although the $D_v$ values of $Bi_2Te_3$/YBCO were comparable to those of BS/YBCO, MT/YBCO exhibited a shorter decay length. Furthermore, different chalcogenides displayed distinct *I-V* characteristics (Supplementary Fig. 14). We find that these curvatures are correlated with both the magnitude of $D_v$ and its dependence on current density. These quantities, in turn, depend on the chalcogenide composition and the spatial distribution of migrated Cu atoms, issues that will be addressed in future work.

**Negative Local Resistance**

Recent experiments in graphene, Weyl semimetals, and other ultraclean metals have stimulated considerable interest in hydrodynamic electron transport in solids.[3,39,40] Among the various experimental signatures, the observation of negative resistance in a vicinity geometry is widely regarded as a hallmark of viscous electron flow.[3,38] The characteristic lateral dimension of such a device is comparable to the hydrodynamic decay length, which is on the sub-mm scale in our samples (Fig. 5a). Accordingly, we fabricated a 1 mm × 2 mm device with appropriately positioned contact pads, as illustrated Fig. 5b.

By generating a highly curved current flow, we observed negative vicinity resistance, consistent with electron backflow or vortex formation. The current dependence of the vicinity resistance was particularly noteworthy. At low temperatures, the measured resistance changed from positive to negative as the current increased (Fig. 5c). The current at which the sign reversal occurred depended on both the vicinity geometry and the positions of the voltage probes.

Previous studies of high-mobility graphene have shown that negative local resistance is indicative of electron backflow and hydrodynamic vortices. Then, the observed transition from positive to negative resistance with increasing current may signal the onset of vortex formation in the electron fluid. Although a quantitative description of this behavior is beyond the scope of the present work, the observation is qualitatively consistent with a hydrodynamic transport regime. The temperature dependences shown in Figs. 5d-5f further reveal that the superconducting transition of the YBCO layer has little effect on the negative local resistance. This result provides additional evidence that the hydrodynamic behavior originates primarily from the chalcogenide layer rather than the superconducting YBCO layer.

## DISCUSSION

The emergence of MNV is not the only anomaly observed in the nonlocal conductor. The pronounced enhancement of the local potential near the current source is equally unusual. At low temperatures, even when the externally applied current flows through a superconducting path, the top surface exhibits exponentially decaying potential profiles in both the current-carrying and current-free regions. As shown by the linearized Navier-Stokes equation, one plausible interpretation is the existence of an electron fluid that propagates in all directions.[41] If this picture is correct, it implies macroscopic conservation of electron momentum over unprecedented length scales. The criterion for the validity of the hydrodynamic regime is $l_{MC} \ll W \ll l_{MR}$ , where $l_{MC}$ is the momentum-conserving electron-electron mean-free path, $W$ is the characteristic system dimension, and $l_{MR}$ is the momentum-relaxation length determined by impurity and phonon scattering. The associated decay length corresponds to the Gurzhi length ($D_v \equiv \sqrt{l_{MC} l_{MR}}$).[39] This criterion has been successfully applied to ultraclean graphene and related systems.[3]

However, the hydrodynamic transport observed in the nonlocal conductor differs markedly from that reported in graphene. First, the extracted Gurzhi length is nearly three orders of magnitude larger than that observed in high-mobility graphene, even at room temperature. This behavior cannot be attributed to the crystalline quality of the chalcogenide layer, because the films investigated here are less crystalline than optimized BS layer with $l_{MR}$ of < 100 nm. Thus, the observations appear to require an unconventional transport regime characterized by an exceptionally large value of the parameter associated with $\boldsymbol{J}_v$. Second, the magnitude of the MNV is extraordinarily large. In graphene, hydrodynamic effects typically appear as subtle corrections to the longitudinal resistivity.[3] On the contrary, in the nonlocal conductor, the local voltage near the current source is dominated by the hydrodynamic contribution over the entire temperature range investigated (Fig. 4b). Third, the assumption of purely two-dimensional transport unlikely to be valid. The measured potential corresponds to the top surface of the bilayer, whereas the potential within the underlying YBCO layer is expected to differ. At low temperatures, when the YBCO becomes superconducting, the potential should be uniform throughout the superconducting layer and approximately equal to the potential measured on the top surface near the center of the local region, where the hydrodynamic contribution is minimal. Accordingly, near the current source, the top and bottom layers may sustain a potential difference approaching 0.5 V at 500 μA, corresponding to the peak height observed in Fig. 4d. Such a vertical potential gradient implies the existence of rotational current flow with an axis lying in the plane of the film, as illustrated schematically in Fig. 4a. This behavior violates several assumptions underlying Eq. (1). In the absence of a more complete theoretical framework, however, we obtain an order-of-magnitude estimate of the viscous Drude conductivity by assuming $D_v \approx$ 200 μm and $\boldsymbol{J}_v \approx \boldsymbol{J}_O$ in Eq. (3). The resulting conductivity, $\sigma_v \approx$ (1 mΩ cm)$^{-1}$, is comparable to that of BS film, suggesting that the observed MNV is at least quantitatively compatible with a hydrodynamic interpretation.

Within a hydrodynamic framework, the highly nonlinear and saturating *I-V* characteristics resemble the shear-thinning behavior of non-Newtonian fluids. The differential resistance decreases with increasing current, analogous to the reduction of viscosity with increasing shear rate in pseudoplastic fluids. Interestingly, we find that the asymptotic decay of the differential resistance is consistent with predictions from holographic hydrodynamics.[42] In that framework, a vanishing differential resistance corresponds to a strongly coupled relativistic limit. Although further theoretical investigation is required, this observation raises the possibility that the nonlocal conductor represents an exotic strongly correlated quantum phase.

If electrons can propagate over macroscopic distances while largely preserving momentum and the measured voltages originate predominantly from hydrodynamic vorticity, such systems may provide a platform for low-dissipation electronic devices. More broadly, these findings suggest the possibility of a new class of electronic functionality based on nonlocal transport phenomena.

## CONCLUSIONS

In summary, the discovery of macroscopic nonlocal voltages reveals an unconventional mode of electron transport that lies beyond the standard description based on local resistivity. Although the microscopic origin of the nonlocal conductor remains to be established, the coexistence of giant nonlocal voltages, anomalous local potentials, millimeter-scale decay lengths, and negative vicinity resistance suggests the emergence of a novel transport regime.

From a materials perspective, the prospects for further development appear favorable. YBCO films are already produced using scalable roll-to-roll technologies,[43] while $Bi_2Se_3$ can be grown not only by molecular beam epitaxy MBE but also by pulsed-laser deposition[44], suitable for industrial supply.

These findings identify nonlocal conductors as a promising system for studying macroscopic electron hydrodynamics and exploring potential low-dissipation electronic functionalities at room temperature.

## EXPERIMENTAL METHODS

**Sample Growth**. BS and MT thin films were grown on 50 nm-thick $YBa_2Cu_3O_{7-\delta}$ (YBCO) films (Ceraco, Germany) deposited on 500 μm-thick *R*-cut sapphire substrates with a $CeO_2$ buffer layer. Prior to chalcogenide growth, the YBCO layer exhibited a $T_C \approx 85$ K. All samples were fabricated using a conventional Veeco 930 MBE system and a home-made MBE system under ultra-high vacuum conditions. Elemental sources (Bi, Se, Mn, and Te) were supplied from solid-source effusion cells (Veeco, USA). For BS growth, the as-received substrate was first heated 260 °C under a Se flux to achieve surface passivation and subsequently heated to 280 °C for film deposition. The Se-assisted growth protocol has previously been shown to enable van der Waals epitaxy of BS even on amorphous $SiO_2$ surfaces. [26] MT films were grown directly at 280 °C without a Te-passivation step. During growth, the beam-equivalent-pressure flux ratios were maintained at Bi:Se=1:15 for BS and Mn:Te=1:15 for MT. The growth rates were 0.22 nm/min for BS and 0.50 nm for MT. Other chalcogenide films were grown using analogous procedures with the corresponding elemental sources.

**Structural Characterization**. BS/YBCO and MT/YBCO bilayers were structurally and compositionally characterized using X-ray diffraction (XRD; Empyrean, Malvern Panalytical, UK) equipped with a 4 kW X-ray source, and high-resolution transmission electron microscopy (HR-TEM; JEM-210F, JEOL, Japan) operated at 200 keV. TEM specimens were prepared by ion-beam milling. Elemental compositions were determined using an energy-dispersive X-ray spectroscopy (EDS) system (Oxford Instruments, UK)

**Transport measurement**. For unpatterned samples, electrical contacts were made using pressed indium contacts. Device patterns were defined by standard photolithography. The chalcogenide layers were etched using Ar plasma, whereas the YBCO layer was chemically etched using 337 mM phosphoric acid. Contact pads were fabricated by photolithography, followed by sequential electron-beam deposition of Cr or Ti (20 nm) and Au (100 nm), and lift-off processing.

Local and nonlocal voltages were measured using digital multimeters (Agilent 34401A), while direct current was supplied by a source-measure unit (Keithley 2400). Temperature-dependent *I*-*V* measurements between 10 and 250 K were performed using a custom-built measurement system integrating a variable-temperature insert with a closed-cycle helium cryostat (Sumitomo (SHI) Cryogenics of America, USA). Measurements at elevated temperatures (260-400 K) were carried out in a controlled-atmosphere glovebox using a custom-built heater stage and a probe station (MS tech, Republic of Korea).

**Raman spectroscopy.** Raman spectroscopic experiments were performed using a 515 nm laser (Cobolt Fandango) with a laser power below 0.3 mW at the sample position. Measurements were carried out at ambient conditions using a home-built microscope stage in backscattering configuration. Inelastically scattered light was dispersed through a Princeton Instruments TriVista 777 spectrometer and recorded with a PyLoN eXcelon charge-coupled device detector.

**Density functional theory**. DFT calculations were performed using the Vienna *Ab initio* Simulation Package (VASP) with the projector augmented wave (PAW) method and the generalized gradient approximation (GGA). A plane-wave cutoff energy of 550 eV and a Γ-centered k-point mesh of 15 × 15 × 2 were used. The structures were relaxed until forces were below 0.01 eV/Å. For $Cu_{0.33}Bi_2Se_3$, the DFT+U method was applied to Cu 3d orbitals with U = 6.5 eV.

**Magnetization measurement**. Temperature-dependent magnetization (*M*-*T*) measurements of BS (20 nm)/YBCO, BS (40 nm)/YBCO, and bare YBCO films were carried out at the Core Research Facilities of Pusan National University using a commercial SQUID magnetometer (MPMS-3, Quantum Design, Inc.). The magnetic moment was recorded in vibrating sample magnetometer (VSM) mode at a vibration frequency of 12.8 Hz with a scan length of 6 mm and an averaging time of 5 s.

ASSOCIATED CONTENT

**Author Contributions**

Conceptualization: JHJ, SHC; Methodology: JHJ, SKJ, HRN, SK, SP, KRC, JSK, HK, GK, SJ, TY, YJS, DW, JK, HN, SHC; Investigation: JHJ, SKJ, HRN, SK, SP, KRC, JSK, HK, GK, SJ, TY, YJS, DW, SHC; Visualization: JHJ, SKJ, HRN; Funding acquisition: JHJ, SP, YJS, HK, SHC; Project administration: SHC; Supervision: SHC; Writing – original draft: SHC; Writing – review & editing: JHJ, SKJ, SHC

ACKNOWLEDGMENT

We thank Kwon Park, Jung Hoon Han, Seyong Kim, Kyoung-Whan Kim and Philip Kim for helpful discussions. We also thank Hunju Lee, Jonghyun Song and Myung-Hwa Jung for experimental supports. We acknowledge the Crystal Bank at Pusan National University for providing valuable samples. ChatGTP 5.5 was used to correct the grammatical errors and misphrasing.

**Funding:** National Research Foundation of Korea (NRF) funded by the Korean government (MSIT), grant no. RS-2021-NR059064 (SHC); NRF funded by the MSIT, grant no. RS-2024-00460233 (JHJ), IITP-ITRC support program funded by the MSIT, grant no. IITP-2026-RS-2024-00437191 (SHC), National Research Council of Science & Technology (NST) grant by the Korea government (MSIT), grant no. GTL24000-000 (SP), NRF funded by the MSIT, grant no.

NRF-2021R1A2C2007141 (YJS), Institute for Basic Science, grant no. IBS-R036-D1 (YJS), KIAS Individual Grant (CG107001) via the Center for Advanced Computation at Korea Institute for Advanced Study (HK).

**Competing interests:** JHJ and SHC are applying a domestic patent on the material for nonlocal voltage and the method of fabrication.

**Data and materials availability:** All data are available in the main text or the supplementary materials. All other data that support the plots within this paper and other findings of this study are available from the corresponding author upon reasonable request.

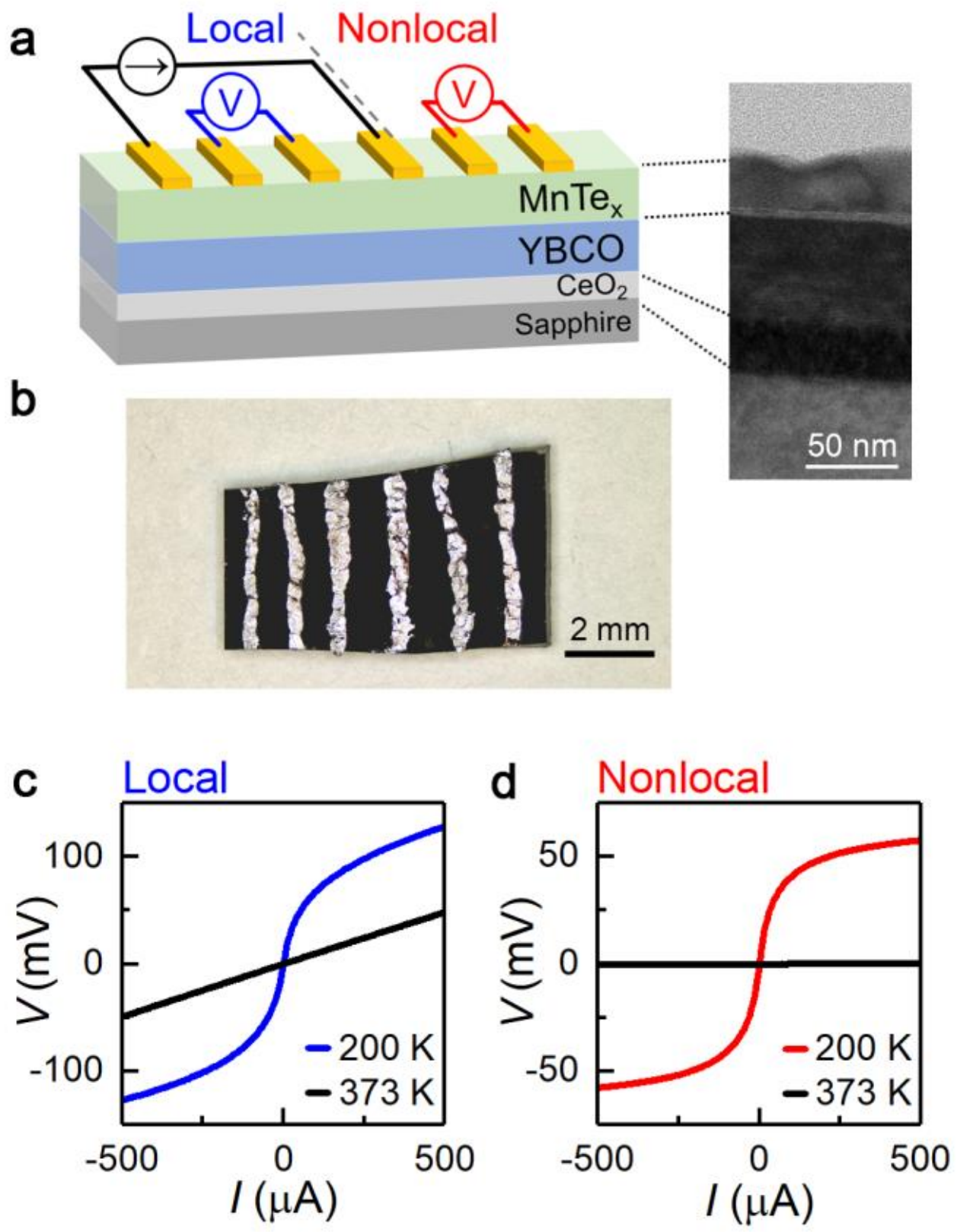

**Fig. 1** Macroscopic nonlocal voltage in MnTe$_x$ (MT)/YBCO. (a) Schematic illustration of the local and nonlocal measurement configurations, together with a cross-sectional TEM image of the sample. (b) Optical image of the sample showing manually applied indium contact lines. (c) Local and (d) nonlocal voltages as functions of the applied current in the local region. Whereas the transport response is Ohmic at 373 K, measurements at 200 K reveal a large macroscopic nonlocal voltage accompanied by a substantial enhancement of the local voltage.

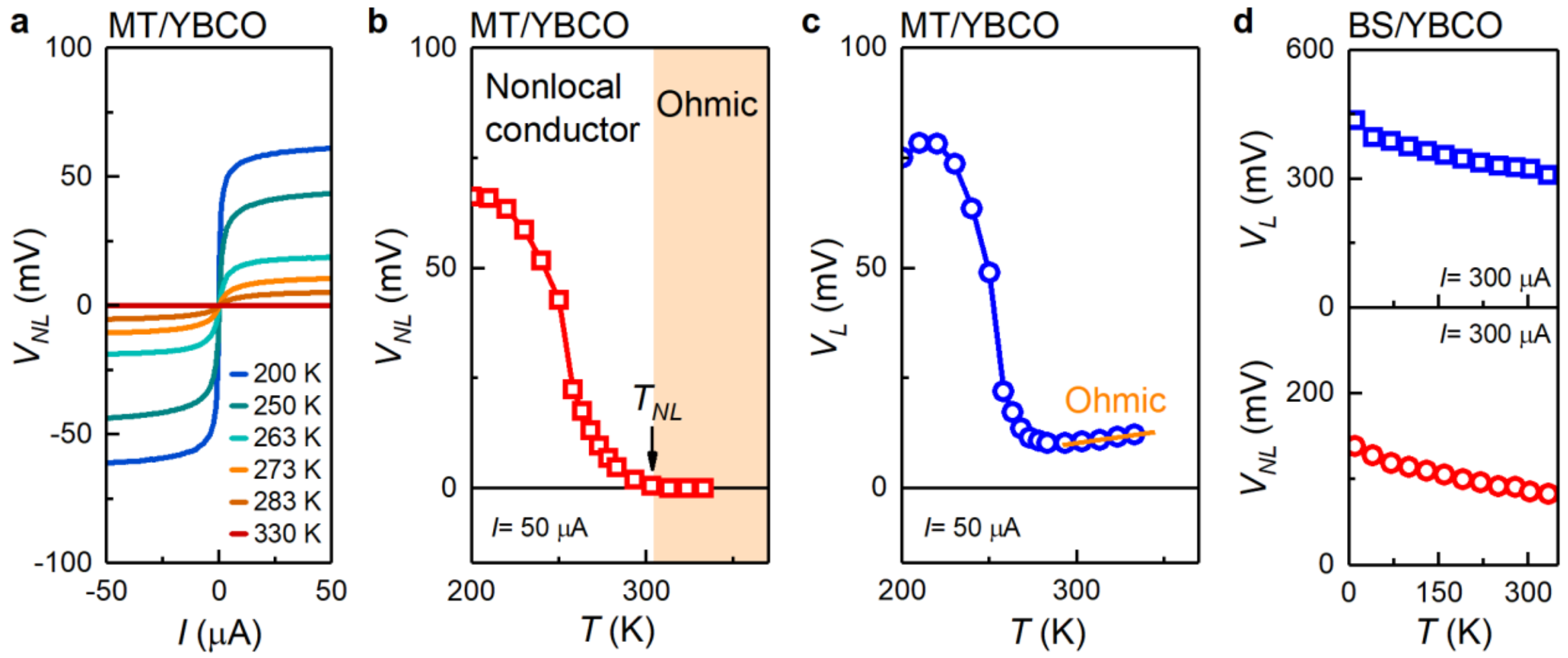

**Fig. 2** Temperature-dependent local and nonlocal voltages of MT/YBCO and $Bi_2Se_3$ (BS)/YBCO. (a) Nonlocal *I*-*V* characteristics of a patterned MT/YBCO sample measured at various temperatures across $T_{NL}$. Temperature dependences of the (b) nonlocal voltage ($V_{NL}$) and (c) local voltage ($V_L$) at $I$ = 50 μA. The disappearance of the nonlocal voltage coincides with the recovery of an Ohmic local response, consistent with a transition from a "nonlocal conductor" to an Ohmic conductor. (d) Temperature dependences of the local and nonlocal voltages in a patterned BS/YBCO sample. In this case, the sample undergoes an irreversible structural change before $T_{NL}$ can be reached experimentally.

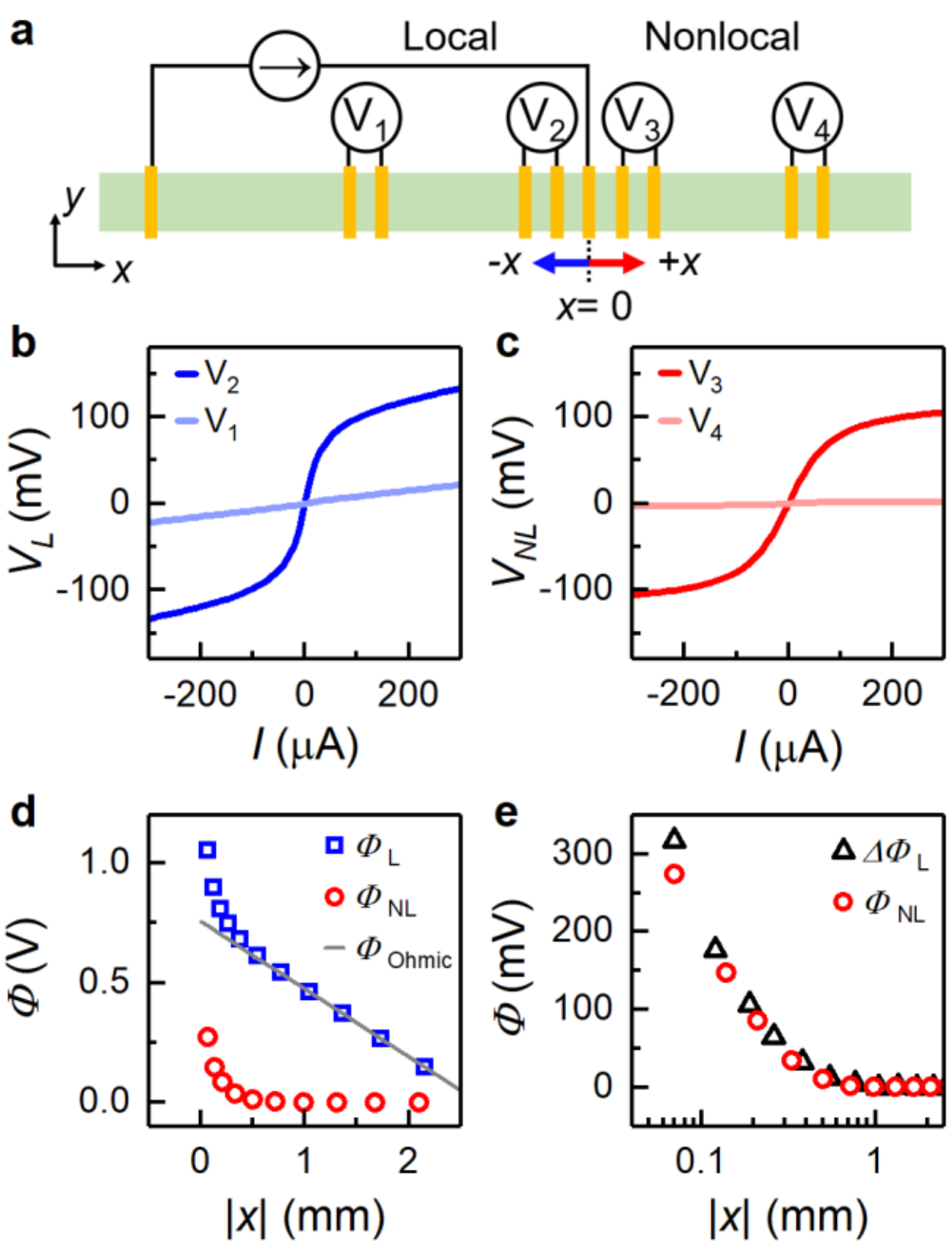

**Fig. 3** Position dependence of local and nonlocal voltages in a patterned BS/YBCO device. (a) Top-view schematic of the measurement geometry. (b) Local ($V_L$) and (c) nonlocal ($V_{NL}$) voltages measured between pairs of voltage probes at $x$ and $x + \Delta x$ ($\Delta x$ = 70 μm for nonlocal and -70 μm for local voltages, respectively). $|x|$ = 70 μm for $V_2$, $V_3$ and 2,550 μm for $V_1$, $V_4$, and $I$ = 300 μA. (d) Potential profiles reconstructed by summing the voltage differences measured between successive probe pairs. The potential at $x$ = -2.5 mm was defined as zero. (e) Comparison of the nonlocal potential ($\phi_{NL}$) and the local potential ($\phi_L$) after subtraction of the Ohmic contribution ($\phi_{Ohmic}$), demonstrating the overlap of the two profiles ($\Delta\phi_L = \phi_L - \phi_{Ohmic}$).

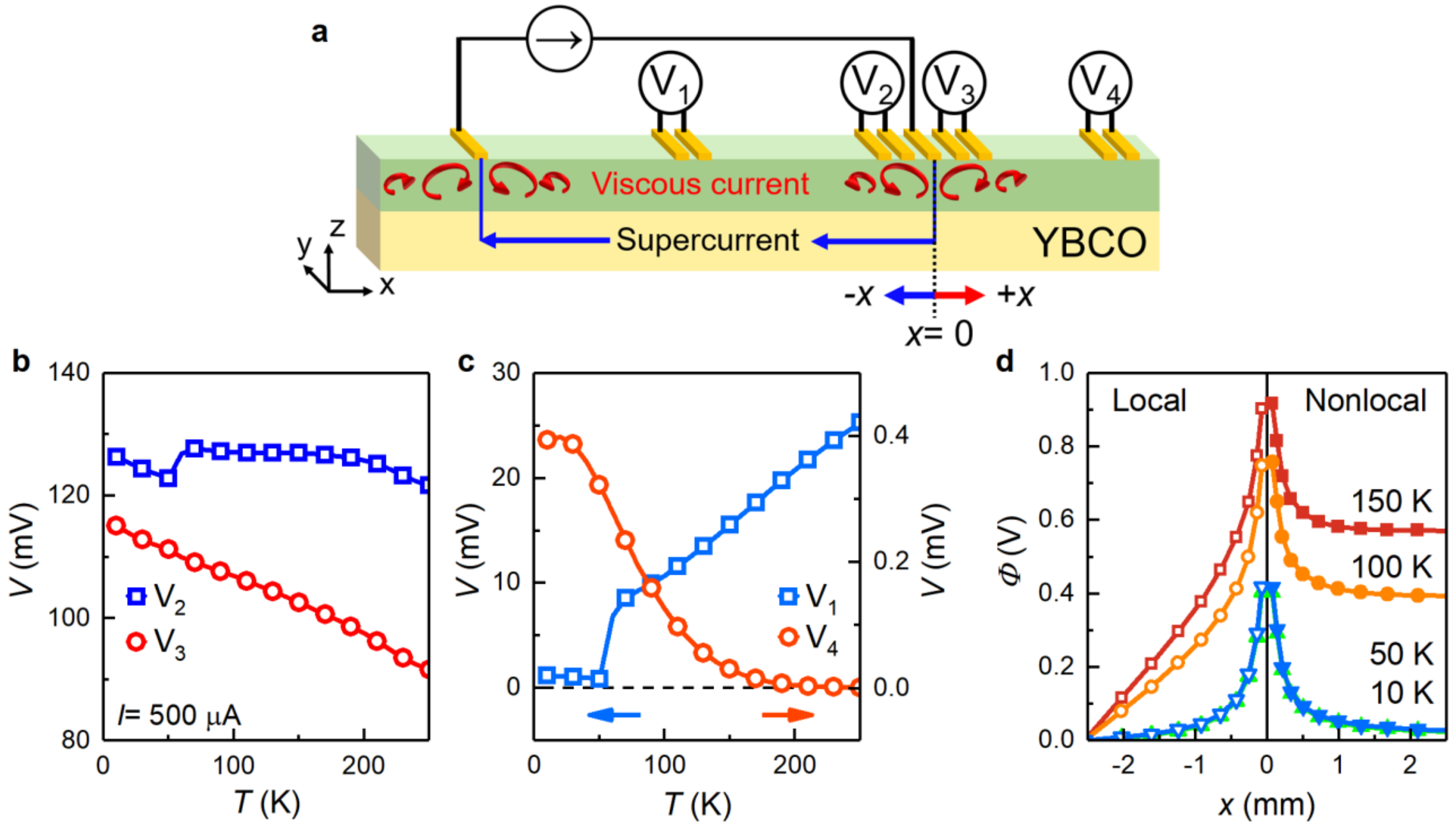

**Fig. 4** Two-current model and temperature-dependent potential profiles in a patterned BS/YBCO device. (a) Schematic illustration of the proposed two-current model used to account for the observed anomalous transport behavior. For simplicity, the YBCO layer is assumed to be in the superconducting state. (b) Temperature dependences of the local and nonlocal voltages for $|x|=$ 70 μm. The non-Ohmic contribution increases as the temperature decreases. (c) Temperature dependences of the local and nonlocal voltages for $|x|=$ 2,550 μm. The local voltage exhibits the characteristic temperature dependence of a superconductor. (d) Potential profiles measured at different temperatures. When the YBCO layer is in the normal state, a symmetric potential peak is superimposed on the conventional Ohmic potential profile. When the YBCO layer becomes superconducting below 50 K, the local and nonlocal potential profiles become nearly identical. $I$ = 500 μA in (b, c, and d).

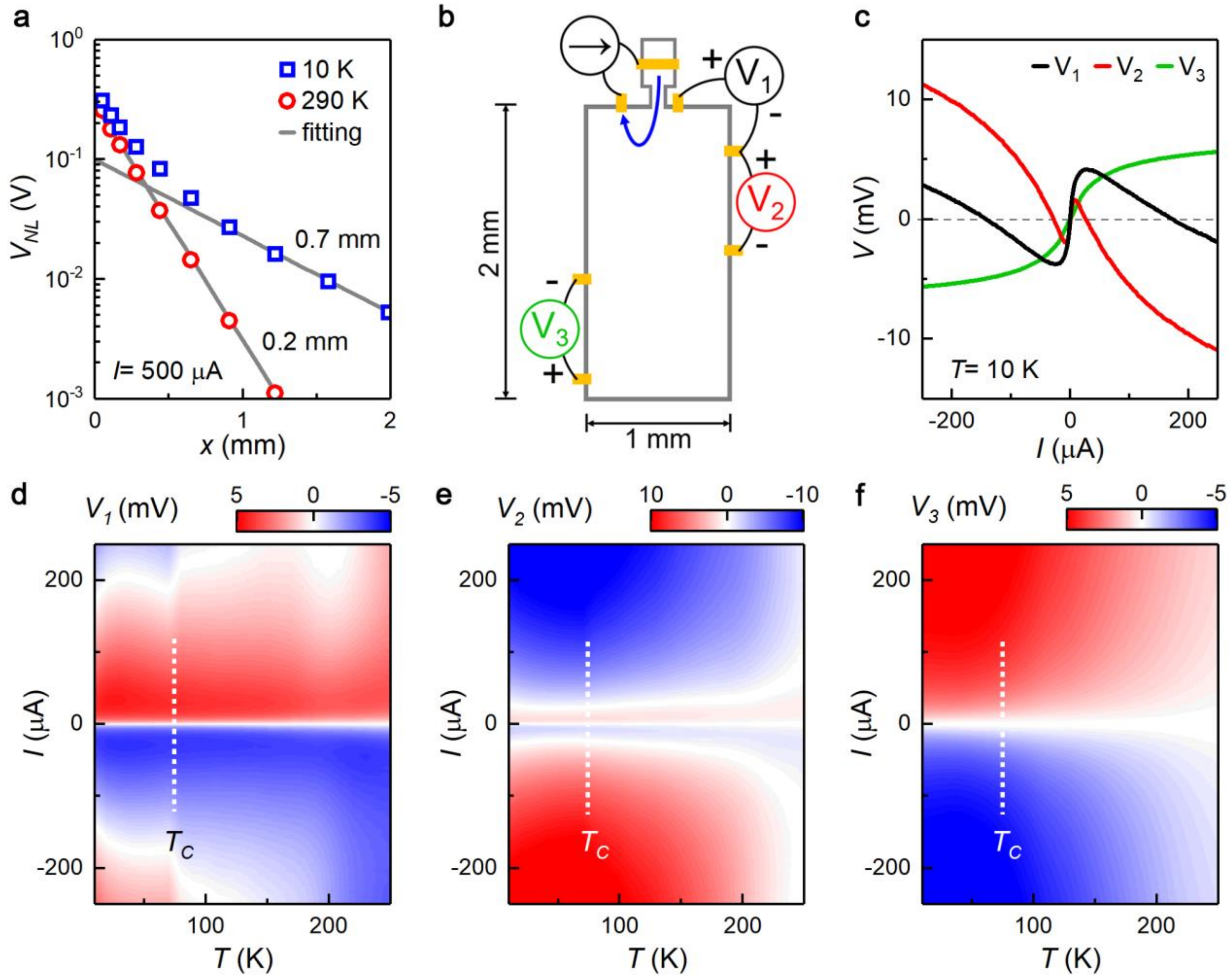

**Fig. 5** Negative local resistance in a vicinity geometry. (a) Nonlocal voltage measured in a patterned BS/YBCO device at 290 K and 10 K, plotted on a logarithmic scale to highlight the exponential spatial decay. The extracted decay length increases from approximately 0.2 mm at 290 K to 0.7 mm at 10 K. (b) Schematic of the macroscopic vicinity device geometry, showing the current path and voltage-probe configuration. (c) Local voltage as a function of current at 10 K. The vicinity contacts ($V_1$ and $V_2$) exhibit a transition from positive to negative voltage as the current increases, resulting in negative local resistance at high currents. (d–f) Local voltages as functions of current and temperature. At sufficiently high currents, $V_1$ and $V_2$ remain negative up to 150 K and 250 K, respectively, whereas $V_3$ exhibits positive local resistance over the entire

temperature range investigated. Notably, no sign reversal is observed at the superconducting transition temperature, $T_C$.

# Supplementary Information

## Observation of Macroscopic Nonlocal Voltage at Room Temperature

Jae Ho Jeon*, Hong Ryeol Na, Sahng-Kyoon Jerng, Seyoung Kwon, Sungkyun Park, Kang Rok Choe, Jun Sung Kim, Heeju Kim, Gunn Kim, Sangmin Ji, Taekeun Yun, Young Jae Song, Dirk Wulferding, Jeong Kim, Hwayong Noh, and Seung-Hyun Chun*

Corresponding authors: jeonjh713@gmail.com and schun@sejong.ac.kr

**The PDF file includes:**

Supplementary Figures S1 to S14 :

Figure S10. Magnetic field dependence of the nonlocal and the local voltages for a patterned device of $Bi_2Se_3$/YBCO

Figure S11. Potential profiles for the local and the nonlocal regions for $Bi_2Se_3$ on sapphire and as-received YBCO

Figure S12. Effect of a selective $Bi_2Se_3$ notching on the macroscopic nonlocal voltage

Figure S13. Decay profiles of the nonlocal voltage for different temperatures and the statistical distribution of decay lengths

Figure S14. Macroscopic nonlocal voltages of $Bi_2Te_3$/YBCO, $Sb_2Se_3$/YBCO, and $CrTe_x$/YBCO

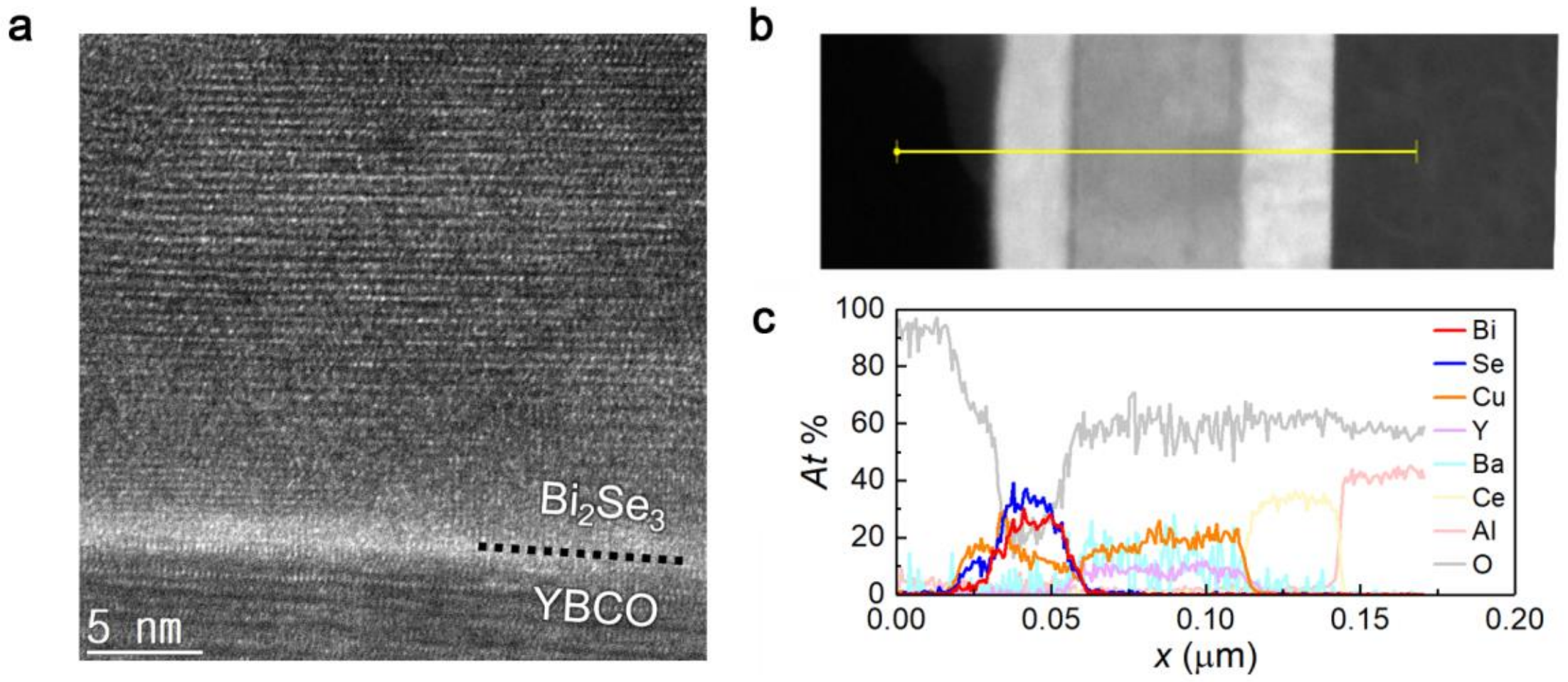

**Figure S1.** High resolution transmission electron microscopy (HR-TEM) and electron dispersive spectroscopy (EDS) of $Bi_2Se_3$/YBCO. (a) Cross-section image shows that c-axis oriented $Bi_2Se_3$ is grown on disordered interface with YBCO. (b), (c) EDS line profile shows a significant Cu diffusion from YBCO into $Bi_2Se_3$.

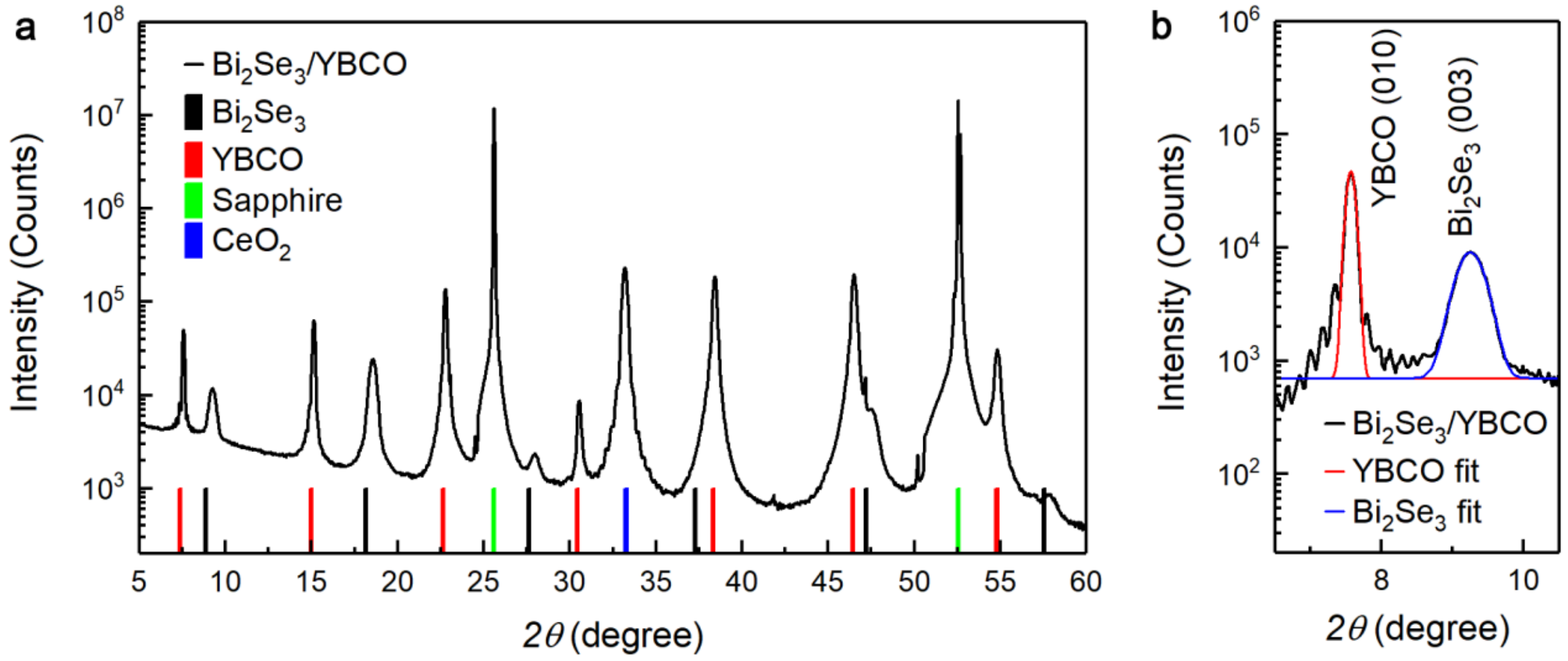

**Figure S2.** X-ray diffraction analysis of $Bi_2Se_3$/YBCO. (a) $2\theta$ scan results are compared with known peaks of $Bi_2Se_3$, YBCO, $CeO_2$, and the sapphire substrate. As all the peaks can be identified, we assume that the Cu atoms were introduced as interstitials. (b) The width of the lowest angle peak was used to extract the average crystallite size of $Bi_2Se_3$ and YBCO to be 19 nm and 84 nm, respectively.

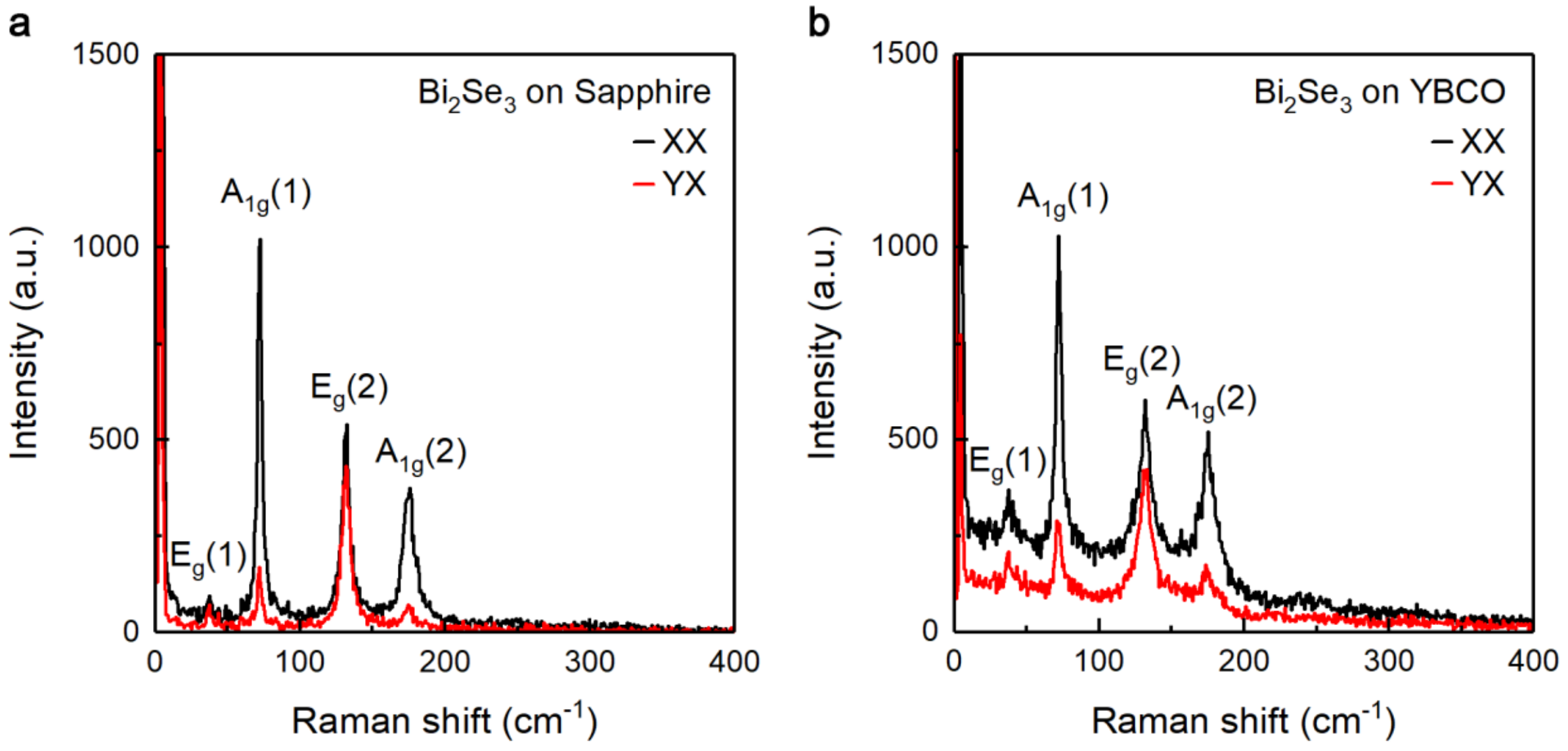

**Figure S3.** Polarized Raman spectroscopy of $Bi_2Se_3$ and $Bi_2Se_3$/YBCO. Raman spectra of $Bi_2Se_3$ on (a) sapphire and (b) YBCO show out-of-plane $A_{1g}$ modes and in-plane $E_g$ modes of $Bi_2Se_3$. Due to the symmetry of $A_{1g}$ modes, the YX polarization makes $A_{1g}$ signals significantly weaker than that of the XX polarization, and other modes involving Cu for $Bi_2Se_3$ are not observed regardless of the polarization, which suggests that diffused Cu atoms from YBCO into $Bi_2Se_3$ were probably introduced as interstitials.

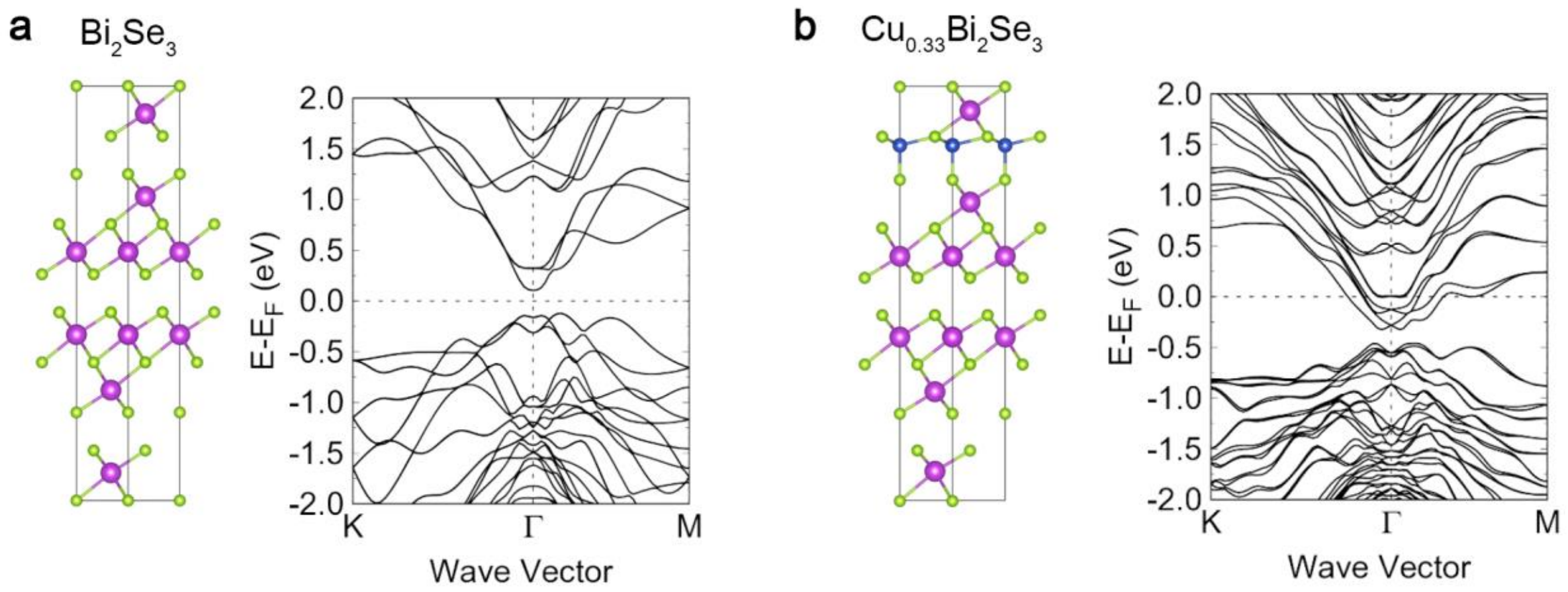

**Figure S4.** Density functional theory calculations of $Bi_2Se_3$ and $Cu_{0.33}Bi_2Se_3$. The optimized structures and the band structures including the spin-orbit coupling are calculated for (a) $Bi_2Se_3$ and (b) $Cu_{0.33}Bi_2Se_3$. Migrated Cu atoms locate within the van der Waals gap of $Bi_2Se_3$ and form tetragonal structures with nearby Se atoms. The elements are color-coded as follows: Bi (magenta), Se (green), and Cu (blue). The lattice constants are almost the same, in accord with Figure S2 and Figure S3 ($a = b = 4.18$ Å, $c = 28.87$ Å for $Bi_2Se_3$ and $a = b = 4.24$ Å, $c = 28.74$ Å for $Cu_{0.33}Bi_2Se_3$).[1]

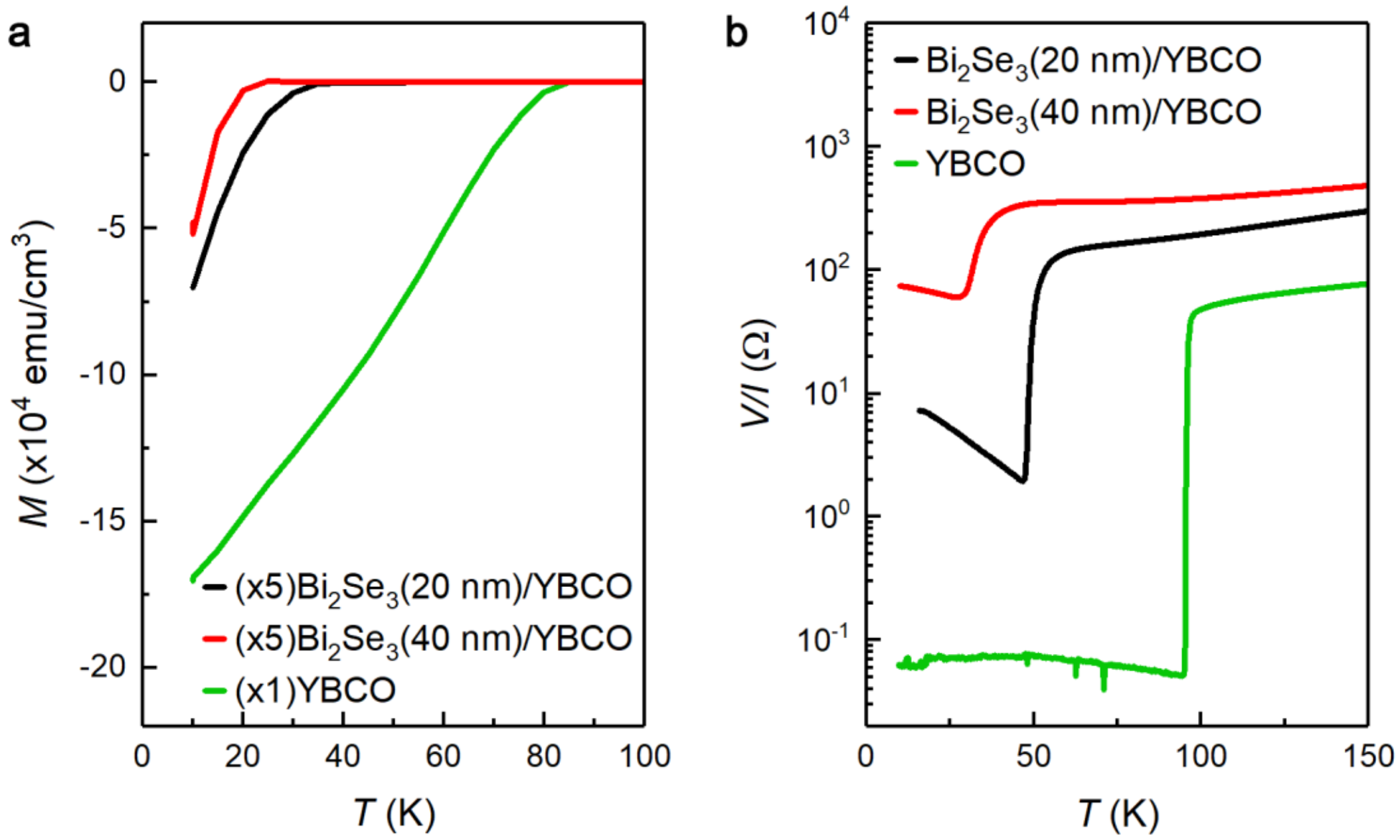

**Figure S5.** Magnetic and electrical characterizations of $Bi_2Se_3$/YBCO and YBCO. (a) Zero-field-cooled magnetization under *H* = 100 G for an as-received YBCO and $Bi_2Se_3$/YBCO bilayers (20 nm-thick and 40 nm-thick $Bi_2Se_3$). As the magnetic field is applied perpendicular to the film plane, the moment reflects the diamagnetic shielding current. (b) High-current *V/I* for the same set of samples as in (a). The current is 300 μA (100 μA for 40 nm-thick $Bi_2Se_3$). Note that the superconducting transition temperatures of YBCO are reduced for $Bi_2Se_3$/YBCO. Finite voltages observed in $Bi_2Se_3$/YBCO at low temperatures originate from the same mechanism giving the macroscopic nonlocal voltages (see the main text on two-current model).

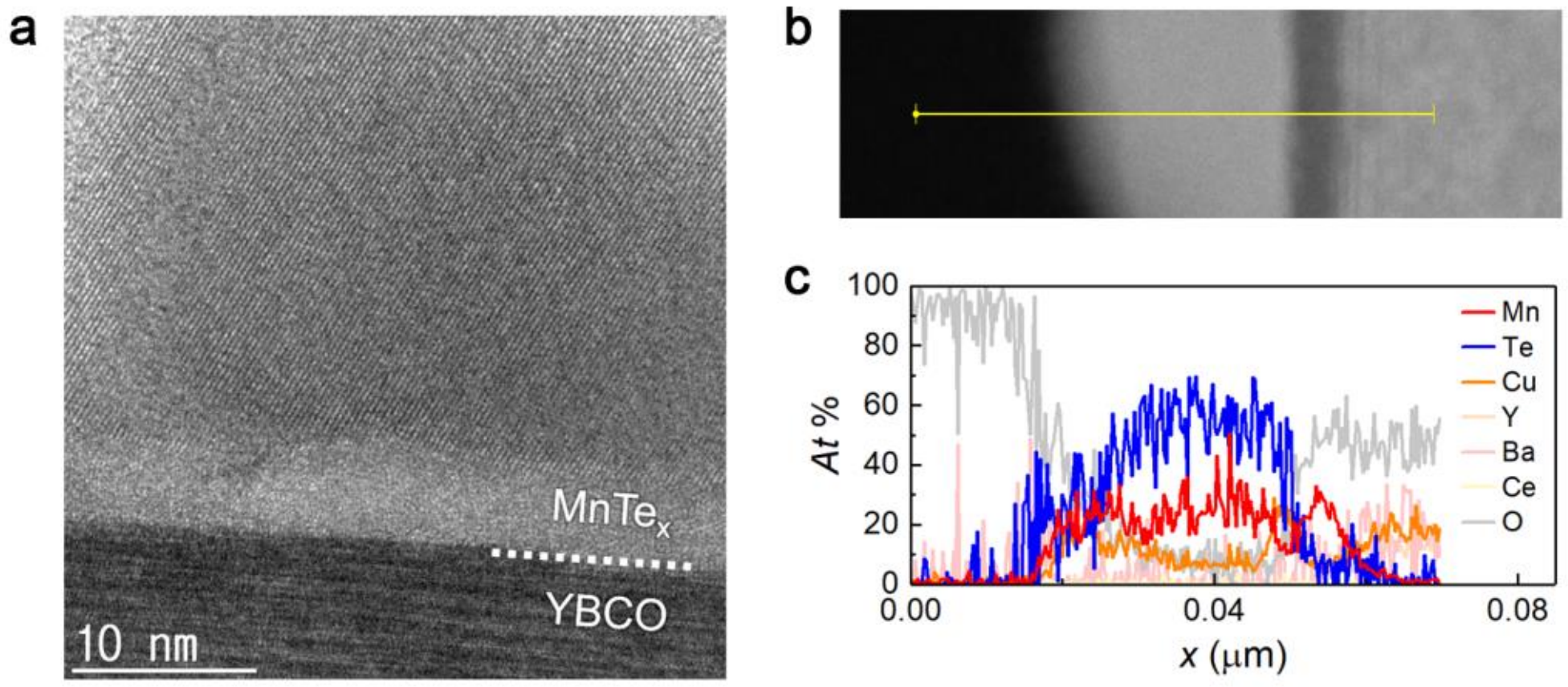

**Figure S6.** HR-TEM and EDS of $MnTe_x$/YBCO. (a) Cross-section image shows that $MnTe_x$ is grown on disordered interface with YBCO. Nanocrystalline $MnTe_x$ is no longer c-axis oriented. (b) (c) EDS line profile shows a significant Cu diffusion from YBCO into $MnTe_x$.

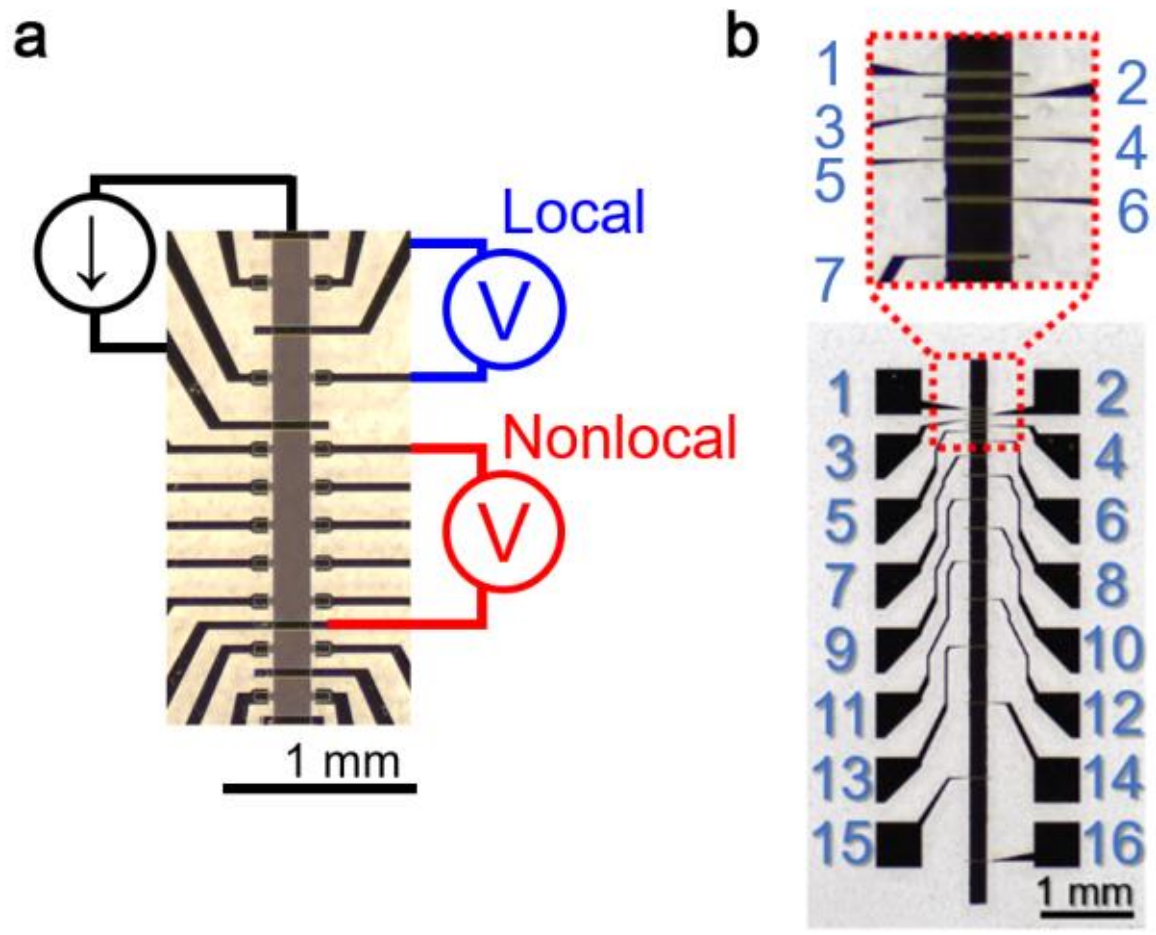

**Figure S7.** Pattern images for Fig. 2 and Fig. 3. (a) A 200 um-width bar is defined by a combination of dry/wet etching. Contacts pad are designed for simultaneous measurements of the local and the nonlocal voltages (Fig. 2). (b) A 200 um-width bar is defined by a combination of dry/wet etching. Contacts pad are designed to measure the local and the nonlocal voltages as a function of distance from the current contact (Fig. 3 and 4).

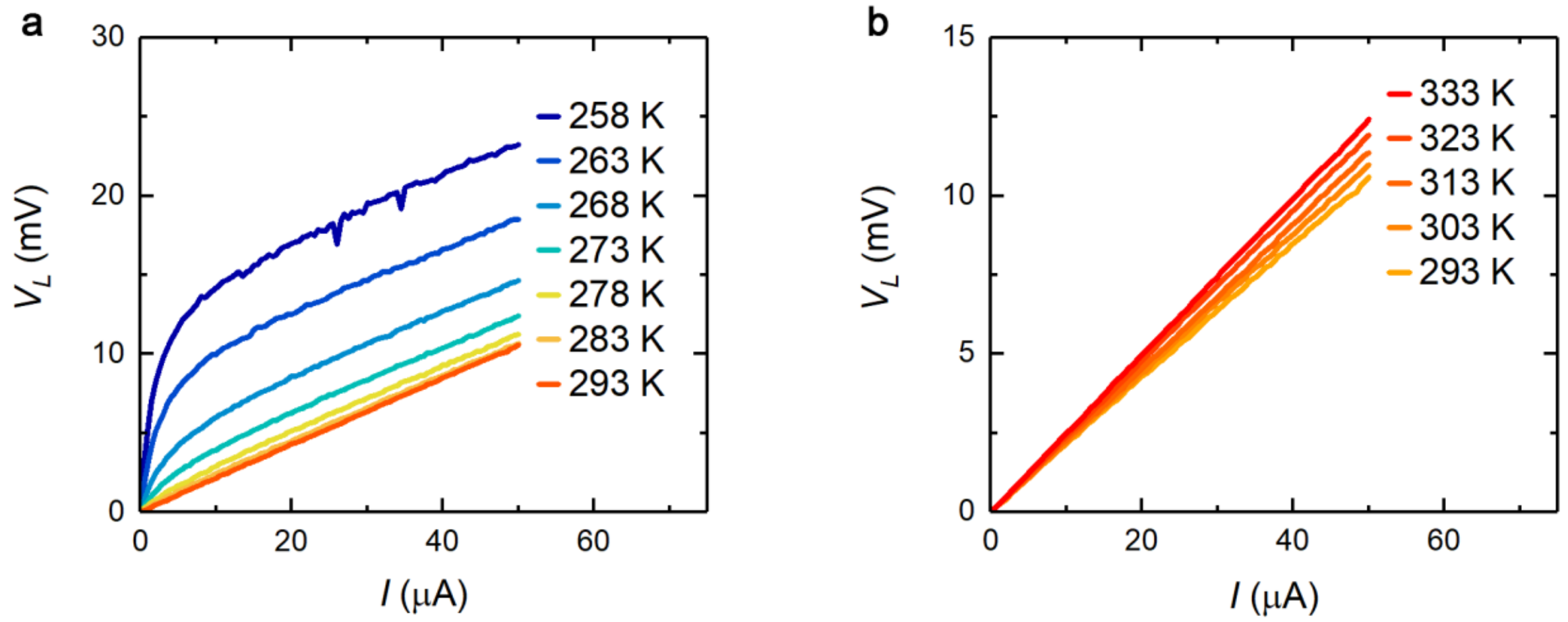

**Figure S8.** Local *I*-*V* characteristics of $MnTe_x$/YBCO for different temperatures. (a) Below the phase transition temperature, *I*-*V* curves are nonlinear. (b)  Near the phase transition temperature, *I*-*V* characteristics restore Ohmic behavior. The slope increases as the temperature increases, as expected for an ordinary metal.

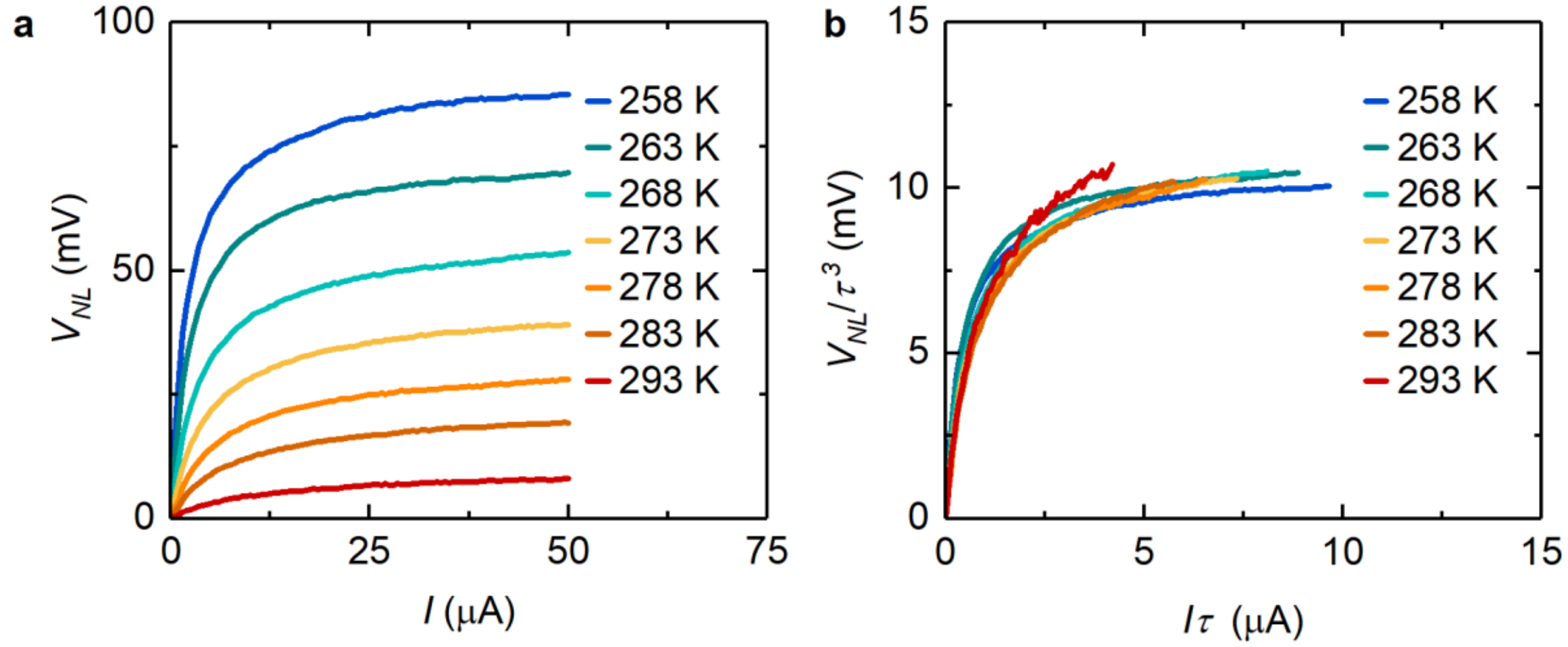

**Figure S9.** Nonlocal *I*-*V* characteristics of $MnTe_x$/YBCO for different temperatures and critical scaling curve. (a) As the phase transition temperature is approached from below, the nonlocal voltage decreases fast. (b) The saturation value of the nonlocal voltage scales as $\tau^3$, where $\tau = 1 - \frac{T}{T_{NL}}$.[2]

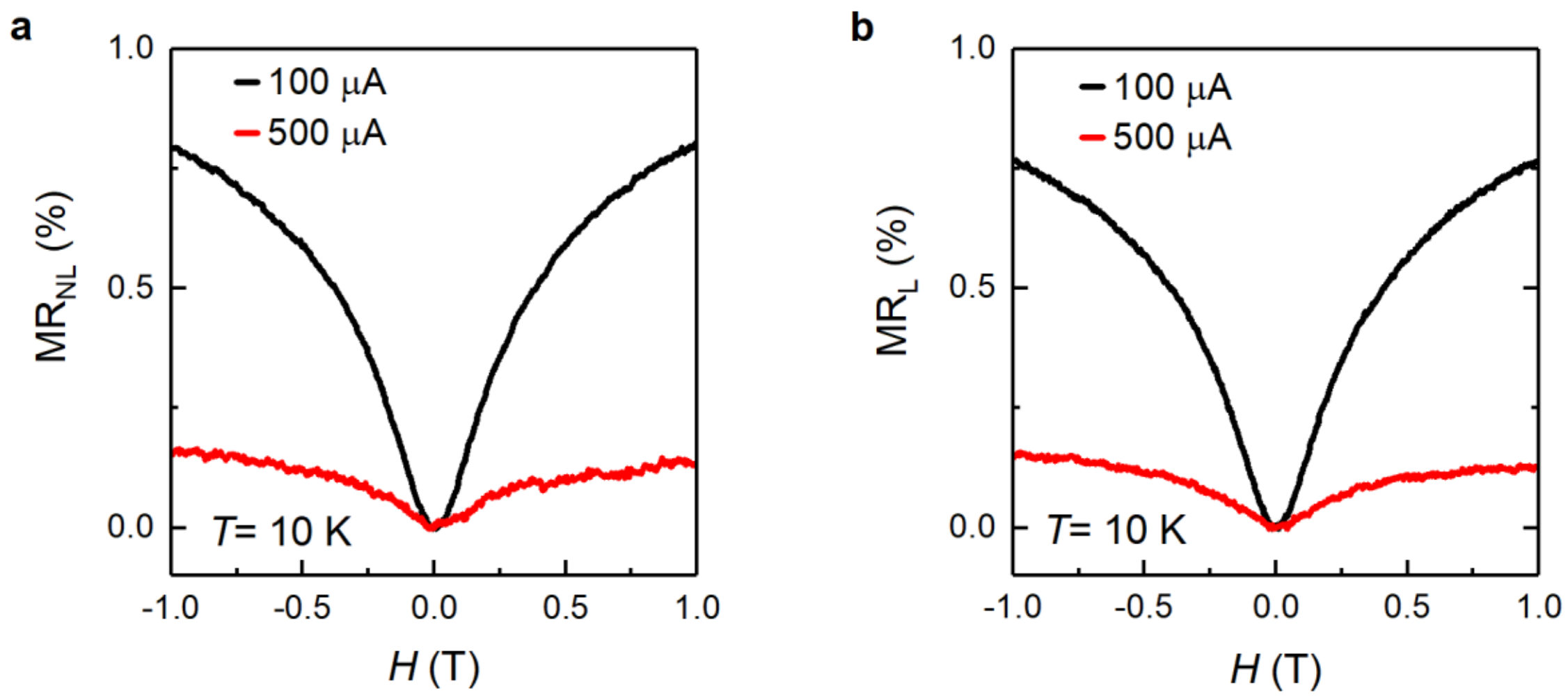

**Figure S10.** Magnetic field dependence of the nonlocal and the local voltages for a patterned device of $Bi_2Se_3$/YBCO. For a perpendicular magnetic field, the changes in the (a) nonlocal and (b) the local voltages are negligible. Here, $\mathrm{MR} \equiv \left(V(H) - V(0)\right)/V(0)$ is a standard formula for magnetoresistance, although the strongly nonlinear and the position-dependent *I*-*V* characteristics make the local resistance ill-defined.

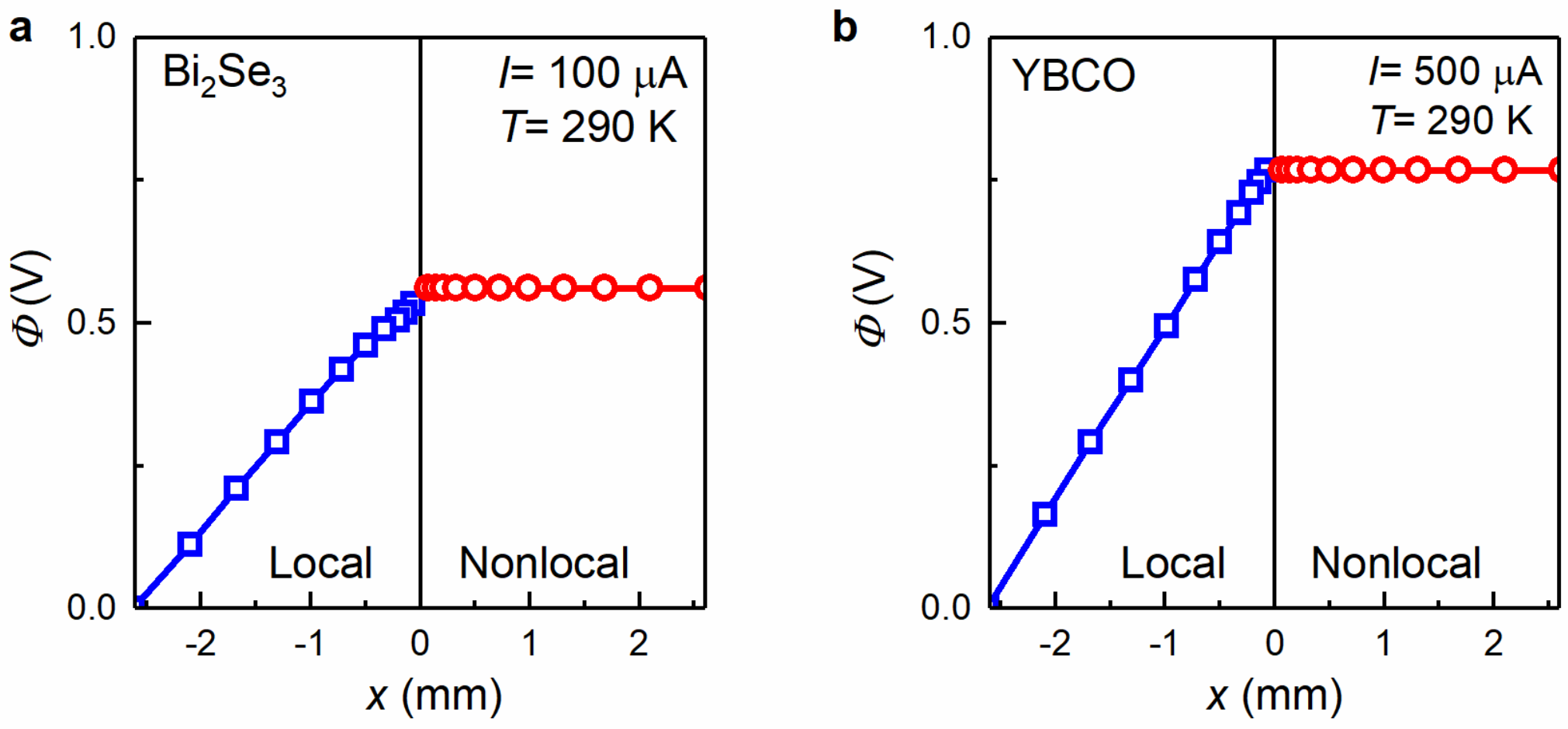

**Figure S11.** Potential profiles for the local and the nonlocal regions for (a) $Bi_2Se_3$ on sapphire and (b) as-received YBCO. For Ohmic conductors, the potential increases linearly in the local region ($x < 0$) and remains constant in the nonlocal region ($x > 0$).

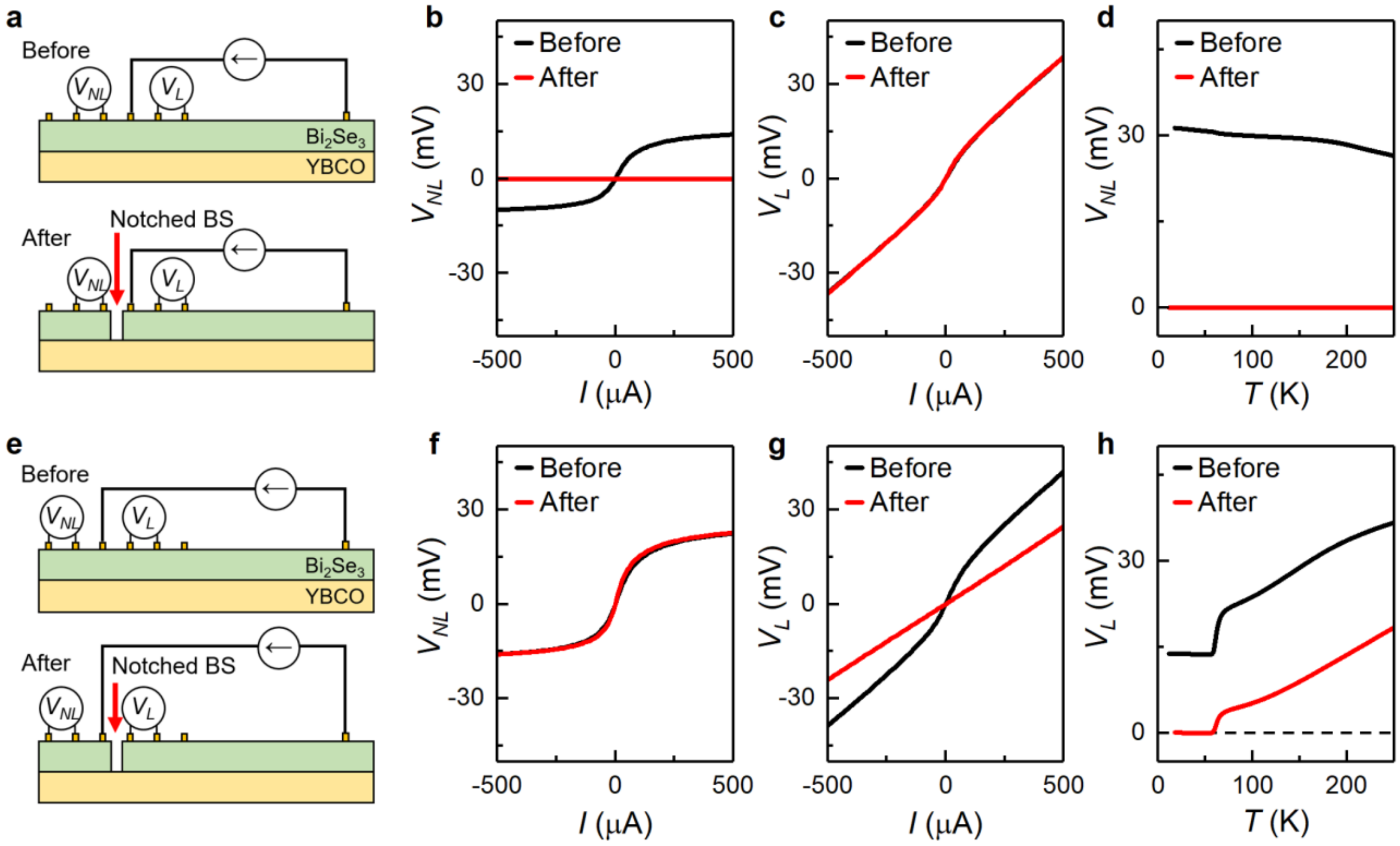

**Figure S12.** Effect of a selective $Bi_2Se_3$ notching on the macroscopic nonlocal voltage. To prove that the nonlocal phase is hosted within the chalcogenide layer, a narrow line of $Bi_2Se_3$ layer was removed across the current direction mechanically. (a) When the notch was made in the nonlocal region, (b) the macroscopic nonlocal voltage disappeared, showing the critical role of chalcogen layer in the appearance of nonlocal voltage. (c) On the other hand, the nonlinear local *I-V* characteristics was preserved. (d) Such behaviors of (b) remained the same for all temperatures. (e) When the notch was made in the local region, (f) the macroscopic nonlocal voltage was unchanged, while (g) the nonlinear local *I-V* characteristics changed to Ohmic. (h) The notched sample showed a typical superconducting transition. (b), (g) imply that the anomalous transport phenomena come from the chalcogen layer.

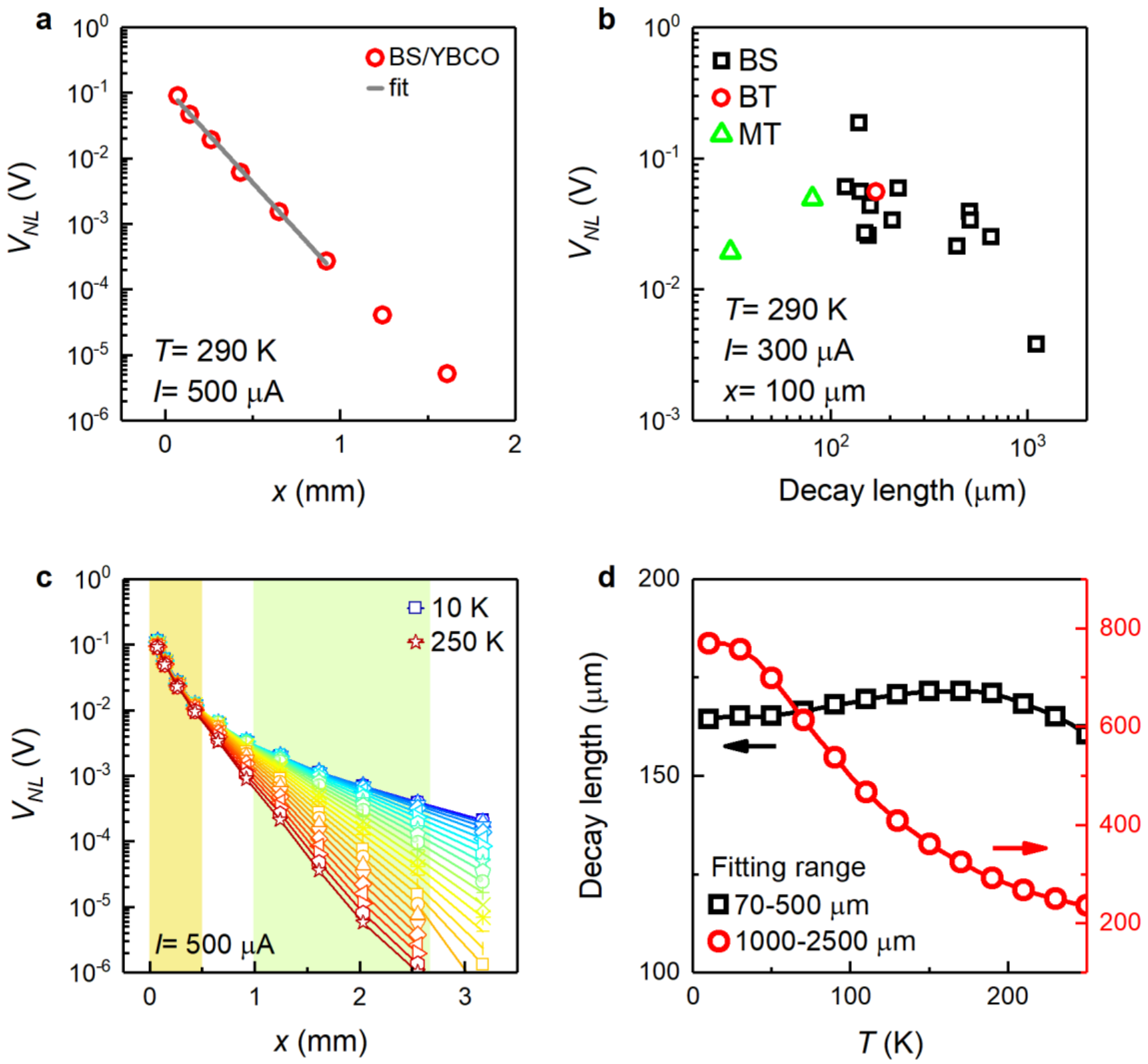

**Figure S13.** Decay profiles of the nonlocal voltage for different temperatures and the statistical distribution of decay lengths. (a) The linear slope in a semi-log plot of nonlocal potential returns the decay length of hydrodynamic signal. (b) Data from patterned samples were collected and compared with the magnitude of macroscopic nonlocal voltage measured at $x$ = 100 μm. BS = $Bi_2Se_3$, BT = $Bi_2Te_3$, and MT = $MnTe_x$. (c) Temperature dependence of the nonlocal potential profile shows that the slope depends on the fitting range. (d) While the decay length remains the same for small $x$, it increases significantly as the temperature decreases for large $x$.

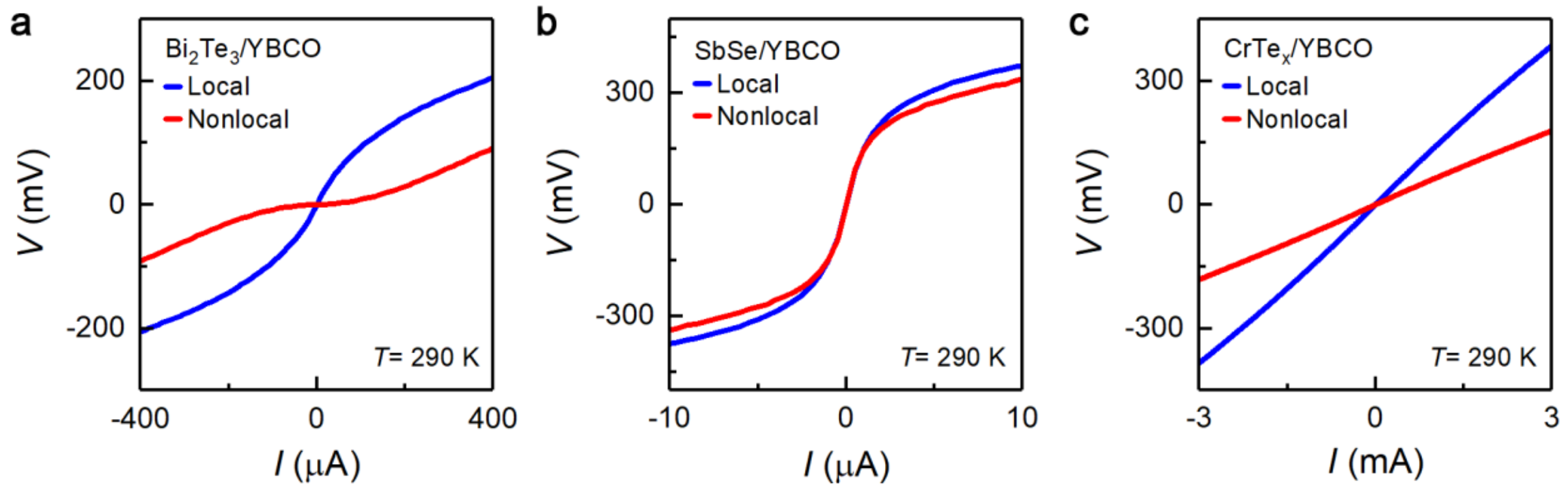

**Figure S14.** Macroscopic nonlocal voltages of $Bi_2Te_3$/YBCO, $Sb_2Se_3$/YBCO, and $CrTe_x$/YBCO. Nonlocal voltages are measured at room temperature for (a) $Bi_2Te_3$/YBCO, (b) $Sb_2Se_3$/YBCO, and (c) $CrTe_x$/YBCO. The sample dimensions are similar to Figure 1b.